\documentclass[12pt]{article}

\usepackage[a4paper,margin=2.2cm]{geometry}
\usepackage[T1]{fontenc}
\usepackage[utf8]{inputenc}
\usepackage{lmodern}
\usepackage{microtype}
\usepackage{amsmath,amssymb,mathtools}
\usepackage{braket}
\usepackage{booktabs}
\usepackage{enumitem}
\usepackage[dvipsnames]{xcolor}
\usepackage{cite}
\usepackage{hyperref}
\usepackage{graphicx}
\usepackage{subcaption}

\hypersetup{
	colorlinks=true,
	linkcolor=blue,
	citecolor=blue,
	urlcolor=blue,
	linktoc=page
}

\numberwithin{equation}{section}
\setlist[itemize]{leftmargin=1.6em,itemsep=0.25em,topsep=0.35em}

\newcommand{\AdS}{\mathrm{AdS}}

\begin{document}

\pagenumbering{gobble}

\title{\textbf{Bulk Criticality and Boundary Spectra in AdS\\
from Matrix Product States}}
\author{Faizan Bhat\thanks{\href{mailto:faizan.bhat@ge.infn.it}{faizan.bhat@ge.infn.it}}\\
\textit{INFN Genova, Via Dodecaneso 33, 16146 Genova, Italy}}
\date{}

\maketitle

\abstract{
	We study interacting scalar quantum field theories in anti-de Sitter (AdS) space using a tensor-network approach based on matrix product states. We develop methods for extracting low-lying boundary spectra from global-AdS energy levels, probing bulk criticality through finite-size scaling in the AdS radius, and identifying conformal boundary conditions. Applying these methods to scalar $\phi^4$ theory in $\mathrm{AdS}_2$, we compute the lowest non-trivial $\mathbb{Z}_2$-odd and even boundary scaling dimensions non-perturbatively. We then locate the bulk $\mathbb{Z}_2$ symmetry-breaking transition and extract critical exponents and central charge consistent with the 2D Ising universality class. At the critical point, we determine the low-lying boundary spectrum and use it to identify the $\mathbb{Z}_2$-preserving free/ordinary Ising conformal boundary condition, thereby characterising both the bulk and boundary universality classes.
}

\clearpage
\tableofcontents
\clearpage
\pagenumbering{arabic}

% -----------------------------------------------------------------------------
\section{Introduction}
% -----------------------------------------------------------------------------

Studying strongly-interacting quantum field theories (QFTs) remains one of the central problems in theoretical physics. In this work, we study interacting QFTs in a fixed anti-de Sitter (AdS) background. As emphasized long ago by
Callan and Wilczek \cite{callan1990INFRAREDBEHAVIOR}, besides being a maximally symmetric spacetime, an advantage of studying QFTs in AdS is that one gains control over the infrared (IR) dynamics  of the theory since the AdS radius $R$ provides a physical IR scale.  AdS also has a conformal boundary on which the bulk theory
defines a natural set of observables which respect
boundary conformal symmetry \cite{aharony2011ConformalField,carmi2019StudyQuantum}.

The boundary description is especially useful because conformal symmetry
remains available even when the bulk theory is massive. Correlation
functions of boundary operators transform under the boundary conformal
group and satisfy the corresponding operator product expansion (OPE) and
crossing equations. This makes it possible to apply conformal-bootstrap
methods even when the bulk theory is away from a conformal fixed point
\cite{paulos2017SmatrixBootstrap,mazac2017AnalyticBounds,
	mazac2019AnalyticFunctional,carmi2019StudyQuantum}.
More generally, boundary conformal data can be used to probe the infrared
dynamics of the bulk theory as the AdS radius is varied. This point of view
has been developed in studies of RG flows, mass-gap generation and
confinement in AdS
\cite{antunes2021BootstrappingRG,meineri2024RenormalizationGroupa,
	lauria2024PerturbativeRGa,copetti2024TamingMass,
	ciccone2024ExploringConfinement,ciccone2026QCDAdS,
	dipietro2026BootstrapStudy},
and has also been applied to supersymmetric gauge theories
\cite{bason2026MathcalN2Super}. In the large-$R$ limit, AdS observables also connect to flat-space
physics. This relation was developed in early work on studying flat-space scattering amplitudes via  the flat-space
limit of AdS/CFT correlators \cite{penedones2011WritingCFT,paulos2017SmatrixBootstrap}, and has since been studied from several complementary points of view
\cite{gary2009LocalBulk,fitzpatrick2011NaturalLanguage,
	raju2012NewRecursion,komatsu2020LandauDiagrams,vanrees2023QuantumField}.
In this sense, varying $R$ provides a continuous interpolation between
curved-space infrared physics and the corresponding flat-space theory.

Another reason to study QFTs in AdS appears when the bulk theory is tuned to a flat-space conformal fixed point. In this case, a CFT on AdS is related by a Weyl transformation to a CFT in flat space with a boundary, and can therefore be viewed as a boundary conformal field theory (BCFT). Different AdS boundary conditions correspond to different conformal boundary conditions, characterized by their boundary spectra and bulk--boundary OPE data
\cite{cardy1984ConformalInvariance,cardy1986EffectBoundary,
	cardy1989BoundaryConditions,cardy1991BulkBoundary,
	mcavity1995ConformalField,liendo2013BootstrapProgram,
	bissi2019AnalyticBootstrap}.
This viewpoint has been applied to interacting scalar, fermionic and gauge theories
\cite{giombi2020CFTAdS,giombi2022FermionsAdS,decesare2026ConformalQED}.  In global AdS, the connection is especially direct, since boundary scaling dimensions are precisely the energy gaps above the vacuum of the bulk theory in units of the AdS radius.

Curvature can also modify infrared phenomena in ways that are interesting in their own right. Studies of confinement and screening in AdS provide recent
examples \cite{aharony2013ConfinementAntide,ankur2026DressingScreening} of this direction of research.
A related question is that of critical phenomena at finite negative curvature \cite{doyon2004IsingField}. Several studies of Ising models on hyperbolic lattices find
mean-field-like critical behavior
\cite{iharagi2010PhaseTransition,gendiar2014MeanfieldUniversality,
	breuckmann2020CriticalProperties},
with similar conclusions in field-theoretic analyses
\cite{benedetti2015CriticalBehavior}.
However, the picture is not completely universal: field-theoretic work has also found non-trivial strong-curvature behavior \cite{mnasri2015CriticalPhenomena}, while studies with frustrated models suggest a richer phase structure \cite{krcmar2008TricriticalPoint}. 

Several complementary approaches have been used to study QFTs in AdS.
At weak coupling, one can work directly in perturbation theory, either in
the bulk or using CFT methods
\cite{bertan2018LoopsSitter,yuan2018LoopsBulk,yuan2018SimplicityAdS,
	bertan2019Quantum$f^4$,aharony2017LoopsAdS}.
Large-$N$ techniques provide another useful analytic and non-perturbative handle
\cite{carmi2019StudyQuantum, dujava2025FinitecouplingSpectrum}. A genuinely non-perturbative numerical approach is Hamiltonian truncation.
Starting from the truncated conformal space approach in two dimensions
\cite{yurov1990TRUNCATEDCONFORMAL}, it has been developed extensively for
interacting QFTs in flat space
\cite{hogervorst2015CheapAlternative,rychkov2015HamiltonianTruncation,
	rychkov2016HamiltonianTruncation,elias-miro2016RenormalizedHamiltonian,
	elias-miro2017HighPrecisionCalculations,elias-miro2017NLORenormalization},
with related developments including lightcone conformal truncation
\cite{katz2016ConformalTruncation,anand2017RGFlow,
	anand2020IntroductionLightcone}. In \cite{hogervorst2021HamiltonianTruncation}, Hamiltonian truncation was formulated directly in AdS$_2$ and applied to interacting theories, including $\phi^4$ theory. This method gives direct access to the low-lying global AdS spectrum and hence to
boundary scaling dimensions. It is naturally geared towards this spectral information, while observables built directly from the bulk states are less immediate to obtain.

In this work, the route we take  instead is based on a lattice Hamiltonian.
Previous lattice studies of QFT in AdS have used regular tessellations of
hyperbolic space, both in AdS$_2$ \cite{brower2021LatticeSetup} and
AdS$_3$ \cite{brower2022HyperbolicLattice}. These lattices preserve a
large discrete subgroup of the AdS isometries, but their microscopic
spacing is naturally tied to the curvature radius. Reaching scales much
smaller than $R$ therefore requires refining the lattice, at which point many of the  symmetries are no longer preserved.
For this reason, rather than starting from a fixed hyperbolic tessellation,  in this  work, we discretize the continuum global AdS$_2$ Hamiltonian directly. This breaks the AdS isometries as well, but allows the lattice spacing and the AdS radius to be varied independently.

We solve the resulting one-dimensional quantum system using a tensor network formulation based on matrix product
states (MPS). MPS and the related density-matrix renormalization group (DMRG) are
standard tools for one-dimensional quantum systems
\cite{white1992DensityMatrix,schollwoeck2011DensitymatrixRenormalization},
and have also been applied directly to relativistic QFTs and lattice gauge
theories
\cite{sugihara2004DensityMatrix,milsted2013MatrixProduct,
	banuls2013MassSpectrum,buyens2014MatrixProduct}. In particular, MPS have been used to study the critical region of
two-dimensional $\phi^4$ theory in flat space \cite{milsted2013MatrixProduct}, cosmological
correlators of interacting $\phi^4$ theory in de Sitter space \cite{basumatary2026CosmologicalCorrelators} and more recently, the Schwinger model in AdS$_2$
\cite{bharadwaj2026ConfinementScreening}.
AdS$_2$ is especially well suited to this approach since a constant-time slice is one-dimensional.  The MPS-based method gives direct access to the ground state as well as to low-lying excited states. We can therefore extract boundary scaling dimensions from
the global energy gaps, while also computing ground-state correlation
functions and entanglement properties that probe the bulk physics directly. 

An important ingredient of our approach is the implementation of finite-size scaling directly in the AdS radius. In flat space, finite-size scaling extracts critical data from the dependence on a finite system size $L$ \cite{fisher1972ScalingTheory,campostrini2014FinitesizeScaling}.
In AdS the corresponding infrared scale is set by $R$, with the physical spacing of the global energy levels scaling as $1/R$.
 Therefore one can use $R$ directly as the finite-size variable. By varying $R$ at fixed renormalized dimensionful couplings, the scaling analysis can be done numerically to extract the bulk critical point and its universal data.

At criticality, the same calculation also has a natural BCFT
interpretation. The global AdS energy gaps become boundary scaling
dimensions, and the resulting boundary spectrum characterizes the
conformal boundary condition reached in the IR. This allows us to identify both the bulk universality class and the corresponding boundary universality class within the same setup.

We develop these methods for a real scalar field with a general local
potential in global AdS$_2$, and apply them to
$\mathbb Z_2$-invariant $\phi^4$ theory with Dirichlet boundary conditions.
We first determine the lowest non-trivial odd and even boundary scaling
dimensions at fixed $R$, finding agreement with perturbation theory
at weak coupling and proceeding to obtain results in the finite coupling regime. We then locate the bulk $\mathbb Z_2$ symmetry-breaking transition through
finite-size scaling in $R$. Binder cumulants, order-parameter scaling and entanglement scaling give critical data consistent with the
2D Ising universality class. Finally, we obtain the boundary spectrum at the independently determined critical point and use it to identify the $\mathbb Z_2$-preserving free/ordinary Ising conformal boundary condition.

The remainder of the paper is organised as follows.
In Sec.~\ref{sec:ads_qft} we review the global AdS Hilbert space, the
bulk--boundary OPE, conditions for well-defined interacting Hamiltonians in AdS, and  conformal boundary conditions at criticality.
In Sec.~\ref{sec:mps} we introduce the lattice Hamiltonian for a general interacting scalar field theory and its MPS
formulation. In Sec.~\ref{sec:phi4} we specialise to scalar $\phi^4$ theory and determine its low-lying boundary spectrum.
Section~\ref{sec:criticality} contains the finite-size-scaling analysis in AdS and the determination of the universal 2D Ising critical data, followed by the identification of the conformal boundary condition at criticality. We conclude with a discussion of open questions and possible extensions in Section~\ref{sec:discussion}.

\section{QFT in AdS}
\label{sec:ads_qft} 
\subsection{Geometry and Hilbert space}
Euclidean AdS$_2$ can be realised as the hyperboloid
\begin{equation}
  -X_0^2+X_1^2+X_2^2=-R^2,
  \qquad
  X_0>0,
\end{equation}
embedded in $\mathbb R^{1,2}$. Introducing global coordinates through
\begin{equation}
  X^\mu(\tau,r)
  =
  {R\over\cos r}
  \bigl(\cosh\tau,\sinh\tau,\sin r\bigr),
\end{equation}
the metric becomes
\begin{equation}
  ds^2
  =
  {R^2\over\cos^2 r}
  \left(d\tau^2+dr^2\right),
  \qquad
  -{\pi\over2}<r<{\pi\over2}.
  \label{eq:global_ads_metric}
\end{equation}
Thus Euclidean global AdS$_2$ is conformal to an infinite strip, with the conformal boundary located at $r=\pm\pi/2$.  

The isometry group of AdS$_2$ is $SO(1,2)\simeq SL(2,\mathbb R)$. The three generators satisfy the following Lie algebra.
\begin{equation}
  [H,P]=P,
  \qquad
  [H,K]=-K,
  \qquad
  [P,K]=-2H.
  \label{eq:sl2_algebra}
\end{equation}
Since the global time coordinate $\tau$ is dimensionless, its generator $H$ is also dimensionless and the corresponding physical Hamiltonian is $H_{\mathrm{phys}}=H/R$.

The Hilbert space can be decomposed into representations of $SL(2,\mathbb R)$. The AdS-invariant vacuum state $\ket{\Omega}$ forms the trivial representation. Choosing the zero of energy such that $H\ket{\Omega}=0$, it is annihilated by all three generators $H$, $P$, and $K$. The non-trivial representations form multiplets with each multiplet consisting of an AdS primary state $|\psi_i\rangle$ that satisfies
\begin{equation}
  H|\psi_i\rangle
  =
  \Delta_i|\psi_i\rangle,
  \qquad
  K|\psi_i\rangle=0 \,,
  \label{eq:ads_primary_state}
\end{equation}
and descendant states obtained by acting repeatedly with $P$.
\begin{equation}
  |\psi_i,n\rangle \sim P^n |\psi_i\rangle, \qquad  H|\psi_i,n\rangle
  =
  (\Delta_i+n)|\psi_i,n\rangle, \quad  
  n=0,1,2,\ldots \,.
  \label{eq:ads_descendants}
\end{equation}
In unitary QFTs, $\Delta_i \ge 0$. In terms of the physical Hamiltonian, $\Delta_i$ measures the energy gap of the AdS primary state $|\psi_i\rangle$ above the vacuum in units of $1/R$. 
\begin{equation}
E_i-E_0=\Delta_i/R\,.
\label{eq:energy_dimension_relation}
\end{equation}
The AdS radius therefore sets the physical infrared scale of the theory effectively acting as a finite-sized box.

\subsection{Bulk-Boundary OPE}
\label{sec:bulk_boundary_OPE}
Given any AdS primary state, one can define boundary primary operators $\mathcal O_i$ via the bulk state – boundary operator map \cite{aharony2011ConformalField,carmi2019StudyQuantum},
\begin{equation}
  |\psi_i\rangle
  =
  \lim_{\tau\rightarrow-\infty}
  e^{-\Delta_i\tau}
  \mathcal O_i(\tau)|\Omega\rangle \,,
  \label{eq:boundary_state_operator_map}
\end{equation}
where the global AdS energy gaps are now interpreted as the scaling dimensions $\Delta_i$ of the boundary primary operators. The same map applied to bulk descendant states defines the boundary descendant operators. Thus it is possible to extract the boundary spectrum of scaling dimensions directly via global AdS energy gaps above the vacuum. Furthermore, any bulk operator admits an expansion in boundary
operators through the bulk--boundary OPE \cite{cardy1991BulkBoundary,mcavity1995ConformalField,liendo2013BootstrapProgram}. Focusing for definiteness on the boundary at $r=\pi/2$, the OPE takes the following schematic form for any bulk operator $\Phi(\tau,r)$.
\begin{equation}
  \Phi(\tau,r)
  =
  \sum_i
  c_{i}\,
  \left(\frac{\pi}{2} - r \right)^{\Delta_i}
  \left[
    \mathcal O_i(\tau)
    +\text{boundary descendants}
  \right].
  \label{eq:bulk_boundary_ope}
\end{equation}
Therefore, the leading near-boundary behaviour of $\Phi$ is governed by the boundary operator of lowest scaling dimension that appears in its OPE. Only boundary operators compatible with the quantum numbers of $\Phi$ and allowed by the symmetries preserved by the boundary condition can appear in its bulk--boundary OPE.  Different admissible boundary conditions therefore lead to different
boundary spectra and bulk--boundary OPE data, even when the bulk dynamics is unchanged.

\subsection{Interacting Hamiltonians in AdS}
\label{sec:int_hams}
The bulk--boundary OPE becomes essential once interactions are included.
Since AdS is locally flat, the local ultraviolet divergences have the
same structure as in flat space and are removed by the usual bulk
renormalisation procedure. In addition, the infinite proper volume of a
constant-time slice can lead to radial divergences in the integrated
Hamiltonian \cite{hogervorst2021HamiltonianTruncation,banados2023BulkRenormalization}. A renormalised interaction density $\mathcal{U}(x)$ enters the Hamiltonian as 
\begin{equation}
U = R^{2} \int_{-\pi/2}^{\pi/2}\!\frac{dr}{\cos^{2}r}\,
\mathcal U(r)\,.
\label{eq:int_density}
\end{equation}
The warp factor $1/\cos^2{r}$ diverges near the boundary and the spatial integral is therefore not automatically finite.  A sufficient condition for convergence is that $\mathcal U(r)$ fall off rapidly enough as $r\to\pi/2$. This translates to a condition on the low-lying boundary primaries appearing in the bulk-boundary OPE of $\mathcal U(r)$. 
Suppose the near-boundary expansion of the renormalised interaction density is
\begin{equation}
\mathcal U(\tau,r)
=
c_{0}\,\mathbf 1
+
\sum_{i\neq\mathrm{id}}
c_{i}\,
(\cos r)^{\Delta_{i}}
\mathcal O_{i}(\tau)
+\cdots.
\label{eq:U_boundary_OPE}
\end{equation}
The identity term $c_{0}\,\mathbf 1$ is the vacuum expectation value
$\langle\mathcal U\rangle$.  It produces a state-independent divergence
in~$U$, which can be removed by the shift
$U\to U-\langle U\rangle$;  in
Hamiltonian language this is equivalent to adding a vacuum-energy counterterm. 

After this subtraction, the leading fall-off comes from the lowest-dimension non-trivial boundary primary with scaling dimension
\begin{equation}
\Delta_*
=
\min_{\substack{i\neq\mathrm{id}\\ c_i\neq 0}}
\Delta_i \,.
\end{equation}
The integrand in \eqref{eq:int_density} therefore behaves as $\left( \frac{\pi}{2}-r \right) ^{\Delta_* -2}$ near the boundary; so the integral only converges provided
\begin{equation}
\boxed{\;\Delta_* > 1\;.}
\label{eq:IR_fin}
\end{equation}
Condition~\eqref{eq:IR_fin} removes infrared divergences associated with the AdS volume. If this condition is not satisfied, additional boundary counterterms
may be required to define the Hamiltonian
\cite{hogervorst2021HamiltonianTruncation,banados2023BulkRenormalization}.

\subsection{Bulk criticality and conformal boundary conditions}
\label{sec:bulk_boundary_criticality}

Let us first recall the standard finite-size-scaling description of a
continuous phase transition
\cite{fisher1972ScalingTheory,campostrini2014FinitesizeScaling}. Away from criticality, a relativistic quantum field
theory has a finite correlation length, set by the inverse of its
lowest bulk mass scale,
\begin{equation}
    \xi \sim \frac{1}{\Delta E_{\rm bulk}}
    \sim \frac{1}{m_{\rm phys}} .
\end{equation}
As a coupling $g$ is tuned towards a continuous transition at $g_c$,
the bulk gap closes and the correlation length diverges as
\begin{equation}
    \xi(g)\sim |g-g_c|^{-\nu},
    \label{eq:correlation_length_scaling}
\end{equation}
where $\nu$ is the correlation-length critical exponent. From the
renormalization-group point of view, tuning to $g_c$ places the theory
on the critical surface, so that its long-distance flow approaches an
IR conformal fixed point.

In a finite flat-space system of linear size $L$, this divergence is
cut off by the system size. At criticality,
\begin{equation}
    \xi(g_c,L)\sim L,
    \qquad
    \Delta E_{\rm bulk}(g_c,L)\sim \frac{1}{L}.
\end{equation}
Close to the transition, the dependence on $g$ and $L$ is controlled by
the scaling variable
\begin{equation}
    x=(g-g_c)L^{1/\nu}.
    \label{eq:flat_fss_variable}
\end{equation}
Thus, for a dimensionless observable $\mathcal Q$,
\begin{equation}
    \mathcal Q(g,L)
    =
    F_{\mathcal Q}\!\left((g-g_c)L^{1/\nu}\right)
    +O(L^{-\omega}),
    \label{eq:dimensionless_fss}
\end{equation}
while the lowest bulk gap satisfies
\begin{equation}
    L\,\Delta E_{\rm bulk}(g,L)
    =
    F_{\Delta}\!\left((g-g_c)L^{1/\nu}\right)
    +O(L^{-\omega}).
    \label{eq:gap_fss_flat}
\end{equation}
Here $\omega>0$ denotes the leading correction-to-scaling exponent
\cite{wegner1972CorrectionsScaling}. Finite-size scaling therefore allows the critical coupling, critical
exponents, and other universal data of the IR fixed point to be
extracted from a sequence of finite systems.

AdS provides a different infrared regulator. Although a constant-time
slice of global AdS has infinite proper length, the curvature radius
$R$ provides a physical infrared scale and the global spectrum is
discrete, with physical level spacings of order $1/R$
\cite{callan1990INFRAREDBEHAVIOR,carmi2019StudyQuantum}.
From the point of view of the bulk RG flow, $R$ therefore plays the
role of an effective system size. Therefore, for a QFT in AdS tuned near its flat-space critical point, one expects
\begin{equation}
    \xi_{\rm AdS}\sim R,
    \qquad
    \Delta E_{\rm bulk}\sim\frac{1}{R},
\end{equation}
or equivalently, the scaling dimension of the corresponding boundary primary operator approaches a finite critical value.

The usual finite-size scaling relations can then be carried over to
AdS by replacing the infrared scale $L$ by $R$. In particular,
\begin{equation}
    \mathcal Q(g,R)
    =
    F_{\mathcal Q}\!\left((g-g_c)R^{1/\nu}\right)
    +O(R^{-\omega}),
    \label{eq:ads_dimensionless_fss}
\end{equation}
and
\begin{equation}
    R\,\Delta E_{\rm bulk}(g,R)
    =
    F_{\Delta}\!\left((g-g_c)R^{1/\nu}\right)
    +O(R^{-\omega}).
    \label{eq:ads_gap_fss}
\end{equation}
The flat-space limit is obtained by taking $R\rightarrow\infty$ while
holding the renormalised dimensionful couplings fixed. Curvature then
becomes irrelevant on any fixed physical length scale. If the theory is
simultaneously tuned to its critical surface, this limit approaches the
flat-space infrared CFT. AdS therefore provides a natural setting for
extracting bulk critical data through \textit{finite-size scaling in
the AdS radius}.

\paragraph{Conformal boundary conditions:}
There is an additional piece of universal information at a bulk
critical point. A CFT on AdS is related by a Weyl transformation to a
CFT in flat space with a boundary and may therefore be viewed as a
boundary conformal field theory
\cite{cardy1984ConformalInvariance,mcavity1995ConformalField,giombi2020CFTAdS}. While the bulk couplings flow towards
the bulk conformal fixed point, microscopic boundary conditions may themselves flow under RG and, at a bulk fixed point and in the absence of an additional boundary scale, are
expected to approach conformal boundary conditions.

A conformal boundary condition specifies how the bulk CFT terminates at
the boundary and is characterised by its spectrum of boundary operators
and the corresponding bulk--boundary OPE data
\cite{cardy1986EffectBoundary,cardy1989BoundaryConditions,cardy1991BulkBoundary}.
Different microscopic boundary conditions can flow to the same
conformal boundary condition and therefore belong to the same
\textit{boundary universality class}; conversely, the same bulk CFT may
admit several distinct conformal boundary conditions.

As discussed in Sec.~\ref{sec:bulk_boundary_OPE}, these boundary
operators appear in the bulk--boundary OPE. In global AdS, their scaling
dimensions are encoded directly in the low-lying energy spectrum, $\Delta_i = R(E_i-E_0)$.
The global-AdS spectrum can therefore be used not only to characterise
the bulk theory, but also to determine which conformal boundary
condition is reached in the infrared. We return to this question
explicitly for the 2D Ising universality class in
Sec.~\ref{sec:criticality}.

\section{Lattice Hamiltonian and tensor-network formulation in AdS}
\label{sec:mps}
In this section, we explain the discretisation of the continuum Hamiltonian on the lattice in global AdS$_2$. For concreteness, we restrict the discussion to a real scalar field with canonical kinetic terms and a general local potential. The Euclidean action in global coordinates can be written as
\begin{equation}
  S_E
  =
  \int d\tau\,dr\,
  \left[
    \frac12(\partial_\tau\phi)^2
    +
    \frac12(\partial_r\phi)^2
    +
    \frac{R^2}{\cos^2 r}\,
    \mathcal U\bigl(\phi\bigr)
  \right].
  \label{eq:scalar_Euclidean_action}
\end{equation}
Here $\mathcal U(\phi)$ denotes a general renormalised interaction density. It may contain a quadratic mass term, polynomial interactions, and the bulk counterterms required by the chosen renormalisation prescription. We also assume that it satisfies the IR finiteness condition given in Eq.~\eqref{eq:IR_fin}. The kinetic terms take the same form as in flat space because the AdS$_2$ metric is conformally flat, while the AdS geometry appears through the warp factor $1/\cos^2 r$ multiplying the potential.

To pass to the Hamiltonian formulation, we analytically continue to
Lorentzian global time, $\tau=it$, for which
\begin{equation}
  S_L
  =
  \int dt\,dr\,
  \left[
    \frac12(\partial_t\phi)^2
    -
    \frac12(\partial_r\phi)^2
    -
    \frac{R^2}{\cos^2 r}\,
    \mathcal U(\phi)
  \right].
  \label{eq:scalar_Lorentzian_action}
\end{equation}
The momentum canonically conjugate to $\phi(t,r)$ is therefore
\begin{equation}
  \Pi(t,r)
  \equiv
  \frac{\partial\mathcal L}
       {\partial(\partial_t\phi)}
  =
  \partial_t\phi(t,r),
  \label{eq:canonical_momentum}
\end{equation}
and obeys the equal-time commutation relation
\begin{equation}
  [\phi(t,r),\Pi(t,r')]
  =
  i\delta(r-r').
  \label{eq:continuum_canonical_commutator}
\end{equation}
The corresponding dimensionless global AdS Hamiltonian is
\begin{equation}
  H
  =
  \int_{-\pi/2}^{\pi/2}dr\,
  \left[
    \frac12\Pi^2(r)
    +
    \frac12\bigl(\partial_r\phi(r)\bigr)^2
    +
    \frac{R^2}{\cos^2 r}\,
    \mathcal U\bigl(\phi(r)\bigr)
  \right].
  \label{eq:continuum_ads_hamiltonian}
\end{equation}

\subsection{Lattice discretization}
\label{sec:lattice_discretization}

The continuum Hamiltonian requires two distinct regulators before it
can be treated numerically. Discretizing the spatial direction provides
an ultraviolet cutoff, while the infinite proper extent of a
constant-time slice of global AdS requires a radial cutoff. In global
conformal coordinates, the latter is reflected in the fact that the
asymptotic boundary lies at $r=\pm\pi/2$, where the warp factor
$1/\cos^2 r$ diverges. We therefore first restrict the spatial slice to
\begin{equation}
  -r_{\max}\leq r\leq r_{\max},
  \qquad
  r_{\max}<\frac{\pi}{2},
  \label{eq:radial_cutoff}
\end{equation}
and recover the full AdS geometry by taking
$r_{\max}\rightarrow\pi/2$. The regulated interval is discretized using $N$ lattice sites at
positions $r_j$. Denoting the separation between neighbouring sites by
\begin{equation}
  \delta_j=r_{j+1}-r_j,
\end{equation}
spatial integrals are approximated according to
\begin{equation}
  \int_{-r_{\max}}^{r_{\max}}dr\,F(r)
  \;\longrightarrow\;
  \sum_{j=1}^{N}w_jF(r_j),
  \label{eq:lattice_quadrature}
\end{equation}
where the weight associated with an interior site is
\begin{equation}
  w_j=\frac{\delta_{j-1}+\delta_j}{2}.
\end{equation}
We keep the site positions general for now to allow the same lattice Hamiltonian to describe different spatial discretizations. 

The lattice fields satisfy the weighted canonical commutation
relations
\begin{equation}
  [\phi_i,\Pi_j]
  =
  \frac{i}{w_i}\delta_{ij}.
\end{equation}
It is convenient to absorb the integration weights into canonically normalized variables,
\begin{equation}
  \widetilde\phi_j
  =
  \sqrt{w_j}\,\phi_j,
  \qquad
  \widetilde\Pi_j
  =
  \sqrt{w_j}\,\Pi_j,
\end{equation}
which obey
\begin{equation}
  [\widetilde\phi_i,\widetilde\Pi_j]
  =
  i\delta_{ij}.
\end{equation}
Approximating the spatial derivative by nearest-neighbour finite
differences then gives us the final form of the Hamiltonian on the lattice
\begin{equation}
  H_{\mathrm{lat}}
  ={}
  \frac12
  \sum_{j=1}^{N}
  \widetilde\Pi_j^{\,2}
  +
  \frac12
  \sum_{j=1}^{N-1}
  \frac{1}{\delta_j}
  \left(
    \frac{\widetilde\phi_{j+1}}{\sqrt{w_{j+1}}}
    -
    \frac{\widetilde\phi_j}{\sqrt{w_j}}
  \right)^2
  +
  \sum_{j=1}^{N}w_j  
  \frac{R^2}{\cos^2 r_j}\,
  \mathcal U\left(
    \frac{\widetilde\phi_j}{\sqrt{w_j}}
  \right).
  \label{eq:lattice_hamiltonian}
\end{equation}
In addition, boundary conditions need to be supplied for the treatment of the endpoint terms and may be specified separately. We now describe two choices of discretization of the global AdS Hamiltonian which are naturally suited to extract the boundary spectra and probe bulk criticality respectively. 

\subsubsection{Global-coordinate discretization and the boundary spectrum}
To determine the low-lying spectrum of the global AdS Hamiltonian, it is natural to discretize the regulated interval uniformly in global coordinates,
\begin{equation}
  r_j
  =
  -r_{\max}+(j-1)a_r,
  \qquad
  a_r=\frac{2r_{\max}}{N-1},
  \qquad
  r_{\max}=\frac{\pi}{2}-\epsilon.
  \label{eq:uniform_global_grid}
\end{equation}
where $j = 1, 2, \ldots, N$. The lattice spacing in proper distances is approximately $a_{\rm proper}(r) \simeq R a_r \sec r$, 
so the physical resolution is finest near the centre and becomes
coarser towards the asymptotic boundaries. For this reason, this discretization is directly adapted to the global Hamiltonian and
its lowest normalizable states, whose wavefunctions vanish near the AdS
boundary and are predominantly supported in the interior.

Both the radial regulator and the finite lattice spacing break continuous AdS isometries. In particular, while global time-translations are preserved, the generators $P$ and $K$ do not act as exact symmetries of the regulated lattice Hamiltonian. The finite-lattice spectrum therefore does not organise exactly into the primary and descendant multiplets of $SL(2,\mathbb R)$ described in Sec.~\ref{sec:ads_qft}.  To obtain the continuum global-AdS spectrum therefore, we need to remove both the regulators. At fixed $R$, this can be done by first taking the continuum limit
\begin{equation}
  a_r\rightarrow0
  \qquad
  (N\rightarrow\infty)
\end{equation}
at fixed radial cutoff $\epsilon$, and subsequently taking
$\epsilon\rightarrow0$. As these limits are taken, the spatial isometries of AdS are restored and the low-lying levels reorganise into $SL(2,\mathbb R)$ representations. The resulting energy gaps may then be interpreted as boundary scaling
dimensions through the global-AdS state--operator correspondence discussed
in Sec.~\ref{sec:bulk_boundary_OPE}.

\subsubsection{Proper-distance discretisation and bulk scaling}
To study bulk observables and critical phenomena, it is useful to work
with a lattice whose resolution is uniform in physical distance. We therefore use a uniform proper-distance lattice 
\begin{equation}
\rho_j = \left(j-\frac{N+1}{2}\right)a_\rho, \qquad  
  \rho
  =
  R\,\operatorname{arcsinh}(\tan r)\,, 
  \label{eq:proper_distance_grid}
\end{equation}
where $ j=1,2,\ldots,N$.  The
radial cutoff is $\rho_{\max}= \frac{N-1}{2}a_\rho$. The corresponding grid in global coordinates is obtained via 
\begin{equation}
r_j = \arctan\left(\sinh\frac{\rho_j}{R}\right)\,.
\end{equation}
At fixed $R$ and $a_\rho$, increasing $N$ simply moves the endpoints
farther into the asymptotic region. In practice, the radial cutoff must be taken to be sufficiently large that observables measured in the
central region converge as $\rho_{\max}/R$ is increased. This effectively removes the infrared regulator.

The proper-distance lattice again breaks the continuous spatial
isometries of AdS, but internal symmetries can be
kept exact. For an even interaction density,
\begin{equation}
  \mathcal U(-\phi)=\mathcal U(\phi),
\end{equation}
the lattice Hamiltonian is invariant under
\begin{equation}
  \widetilde\phi_j\longrightarrow-\widetilde\phi_j,
  \qquad
  \widetilde\Pi_j\longrightarrow-\widetilde\Pi_j,
\end{equation}
and the Hilbert space separates into even and odd $\mathbb Z_2$
sectors. This makes the discretization well suited to studying a
transition driven by spontaneous breaking of the same symmetry, since
the regulator does not explicitly favour either phase. For a finite-size scaling analysis as discussed in Sec. ~\ref{sec:bulk_boundary_criticality}, the scaling limit is given by keeping $a_\rho$ fixed and increasing $R$.
\begin{equation}
	\frac{R}{a_\rho}\rightarrow\infty .
\end{equation}
Near a continuous transition, lattice effects in this limit are expected
to become irrelevant and the universal long-distance scaling behaviour
is recovered \cite{fisher1972ScalingTheory,campostrini2014FinitesizeScaling}.
The critical values of the bare couplings remain regulator dependent,
while critical exponents, scaling functions, and boundary scaling
dimensions approach their universal continuum values.

Finally, let us mention that both the discretizations discussed here differ from the hyperbolic-lattice constructions of Refs.~\cite{brower2021LatticeSetup,brower2022HyperbolicLattice}, since
the continuum global coordinate is discretized directly and the
lattice spacing can be varied independently of the AdS radius.

\subsection{Local Hilbert space and basis truncation}
\label{sec:local_hilbert_space}

Once the spatial direction has been discretized, the field theory
becomes a one-dimensional quantum system with one bosonic degree of
freedom at each dynamical lattice site. Before any truncation is made,
the Hilbert space of the regulated theory is therefore
\begin{equation}
  \mathcal H_{\mathrm{lat}}
  =
  \bigotimes_{j=1}^{N}\mathcal H_j,
  \qquad
  \mathcal H_j\simeq L^2(\mathbb R),
  \label{eq:full_lattice_hilbert_space}
\end{equation}
where $N$ is the number of lattice sites. Each local
Hilbert space is infinite dimensional. To construct a discrete local basis, we introduce oscillator operators
at each site,
\begin{equation}
  \widetilde\phi_j
  =
  \frac{a_j+a_j^\dagger}{\sqrt{2\omega_j}},
  \qquad
  \widetilde\Pi_j
  =
  -i\sqrt{\frac{\omega_j}{2}}
  \left(a_j-a_j^\dagger\right),
  \label{eq:local_oscillator_basis}
\end{equation}
with
\begin{equation}
  [a_i,a_j^\dagger]=\delta_{ij}.
\end{equation}
For any positive choice of $\omega_j$, the local Fock states
\begin{equation}
  |n_j\rangle
  =
  \frac{(a_j^\dagger)^{n_j}}{\sqrt{n_j!}}|0_j\rangle,
  \qquad
  n_j=0,1,2,\ldots,
\end{equation}
form a complete basis of $\mathcal H_j$. For numerical calculations, the Fock space at each site is truncated to
\begin{equation}
\mathcal H_j^{(d)}= \operatorname{span} \left\{ |0_j\rangle,\ldots,|d-1_j\rangle \right\}.
\label{eq:local_fock_truncation}
\end{equation}
The resulting finite-dimensional lattice Hilbert space is \begin{equation}
\mathcal H_{\mathrm{lat}}^{(d)} =
\bigotimes_{j=1}^{N}
\mathcal H_j^{(d)},
\qquad
\dim\mathcal H_{\mathrm{lat}}^{(d)} =
d^{N}.
\label{eq:truncated_full_hilbert_space}
\end{equation}
If the complete Fock space were retained, different positive choices of $\omega_j$ would simply correspond to different bases of the same local Hilbert space. After truncation, however, they need not represent the low-energy eigenstates of the full interacting lattice Hamiltonian with the same accuracy at fixed $d$. The choice of local oscillator scale therefore becomes an important part of the numerical formulation.

This is especially important in AdS because the local terms in the lattice Hamiltonian vary strongly across the spatial slice, especially towards the regulated boundary. The interaction density enters with the position-dependent factor $R^2w_j\sec^2 r_j$, while the lattice weights and the diagonal contributions from the gradient term are also site-dependent. A basis defined using a single position-independent oscillator frequency may therefore represent these fluctuations well in one region of the lattice but require many occupation states in another. Allowing the basis frequency $\omega_j$ to vary with position adapts
the local oscillator wavefunctions to the different field scales across the lattice and can substantially reduce the local dimension $d$ required for converged energies and observables. We therefore propose two natural prescriptions for choosing these local frequencies.
\begin{enumerate}
    \item  \textbf{On-site quadratic basis:} A simple choice is obtained by isolating the quadratic part of the terms acting locally at each site. Writing it as
\begin{equation}
\label{eq:onsitebasis}
H_{j,\mathrm{loc}}^{(2)} = \frac12\widetilde\Pi_j^{\,2} +
\frac12\Omega_j^2\widetilde\phi_j^{\,2},
\end{equation}
we choose
\begin{equation}
\omega_j=\Omega_j.
\label{eq:onsite_frequency_choice}
\end{equation}
The local quadratic contribution is then diagonal in the occupation number of the corresponding oscillator. This gives a simple position-dependent basis constructed directly from the local quadratic terms of the lattice Hamiltonian.

\item \textbf{Free-vacuum basis:} When the quadratic part of the lattice Hamiltonian can be diagonalized
exactly, its ground state provides a natural choice of local basis. Denoting the ground state by $|\Omega_0\rangle$, we choose the oscillator frequency at site $j$ as
\begin{equation}
  \omega_j
  =
  \sqrt{
    \frac{
      \langle\Omega_0|
      \widetilde\Pi_j^{\,2}
      |\Omega_0\rangle
    }{
      \langle\Omega_0|
      \widetilde\phi_j^{\,2}
      |\Omega_0\rangle
    }
  }.
  \label{eq:free_vacuum_frequency}
\end{equation}
This matches the local oscillator basis to the field and momentum
fluctuations of the free vacuum, including the effects of the couplings between neighbouring sites. It is useful when the free theory provides a good reference for the low-energy states of the full interacting Hamiltonian.
\end{enumerate}

\subsection{Matrix Product States}
\label{sec:mps_approximation}

Truncating each local Fock space to dimension $d$ makes the Hilbert
space at every lattice site finite dimensional. The dimension of the
full lattice Hilbert space, however, still grows exponentially. For
$N$ dynamical sites, a general state takes the form
\begin{equation}
  |\Psi\rangle
  =
  \sum_{n_1,\ldots,n_N=0}^{d-1}
  C_{n_1\cdots n_N}
  |n_1,\ldots,n_N\rangle ,
  \label{eq:general_lattice_state}
\end{equation}
and is specified by $d^N$ coefficients. Keeping and optimizing all of
these coefficients quickly becomes impractical as the number of sites
is increased.

A matrix product state provides a compact representation of the same
many-body wavefunction by factorizing the coefficient tensor into a
sequence of local tensors \cite{schollwoeck2011DensitymatrixRenormalization},
\begin{equation}
  C_{n_1\cdots n_N}
  =
  \sum_{\alpha_1,\ldots,\alpha_{N-1}}
  A^{[1]\,n_1}_{\alpha_1}
  A^{[2]\,n_2}_{\alpha_1\alpha_2}
  \cdots
  A^{[N]\,n_N}_{\alpha_{N-1}} .
  \label{eq:mps_coefficients}
\end{equation}
The physical index $n_j=0,\ldots,d-1$ labels the local Fock states at
site $j$. The auxiliary index $\alpha_j$, shared by the tensors at
sites $j$ and $j+1$, has dimension $\chi_j$. We denote the largest of
these dimensions by
\begin{equation}
  \chi=\max_j\chi_j ,
\end{equation}
which is called the \textit{bond dimension}. At fixed $d$ and $\chi$, the number
of parameters required to describe the state grows approximately as
\begin{equation}
  N d\chi^2,
\end{equation}
rather than exponentially with $N$.

How large $\chi$ must be to represent a state accurately is determined
by the entanglement of the state. To see this, let us cut the lattice
between sites $j$ and $j+1$. The state can then be written in its
Schmidt decomposition,
\begin{equation}
  |\Psi\rangle
  =
  \sum_{\alpha=1}^{r_j}
  s_\alpha^{(j)}
  |\alpha\rangle_{\mathrm L}
  \otimes
  |\alpha\rangle_{\mathrm R},
  \qquad
  \sum_{\alpha=1}^{r_j}
  \left(s_\alpha^{(j)}\right)^2=1 ,
  \label{eq:schmidt_decomposition}
\end{equation}
where $r_j$ is the Schmidt rank across the cut. The corresponding
entanglement entropy is
\begin{equation}
  S_j
  =
  -\sum_{\alpha=1}^{r_j}
  \left(s_\alpha^{(j)}\right)^2
  \log\left(s_\alpha^{(j)}\right)^2 .
  \label{eq:entanglement_entropy}
\end{equation}

An MPS with bond dimension $\chi_j$ has Schmidt rank at most $\chi_j$
across this cut. It therefore satisfies
\begin{equation}
  S_j\leq\log\chi_j .
  \label{eq:mps_entropy_bound}
\end{equation}
The bond dimension thus directly controls how much entanglement the MPS
can represent between the two parts of the lattice. A state with larger
entanglement requires a larger value of $\chi_j$.

Of course, any state in the finite lattice Hilbert space can be written exactly as an MPS if the bond dimensions are allowed to become large
enough. Exact representation requires $\chi_j=r_j$
at every cut and since $r_j \leq \min\left(d^j,d^{N-j}\right)$, a general state requires $\chi\sim d^{N/2}$, and the exponential complexity of the full Hilbert space is then recovered.

The MPS approximation is therefore useful for states whose entanglement remains sufficiently small as the lattice size is increased. If the entropy approaches a constant, a fixed bond
dimension can in principle remain sufficient. If instead $S_j\sim\log N$, it is still manageable as  then, Eq.~\eqref{eq:mps_entropy_bound} implies that the required bond dimension grows only as a power of $N$. In the next subsection, we show that for the extraction of continuum boundary spectra and for probing criticality in AdS, the entanglement required  grows only logarithmically with the relevant system size, making these problems well suited to an MPS description.

\paragraph{Obtaining ground and excited states.}
The ground state $|\Psi_0\rangle$ is obtained by varying the MPS coefficients in \eqref{eq:mps_coefficients} (sometimes called the MPS manifold) to minimize the energy
\begin{equation}
	E_0
	=
	\min_{\Psi}
	\frac{\langle\Psi|H_{\mathrm{lat}}|\Psi\rangle}
	{\langle\Psi|\Psi\rangle}.
	\label{eq:mps_variational_energy}
\end{equation}

When the lattice Hamiltonian has a $\mathbb Z_2$ symmetry
$\phi\mapsto-\phi$, we denote the corresponding parity operator by
$\hat P$, with eigenvalues $P=+1$ and $P=-1$ in the even and odd sectors,
respectively. Rather than imposing the $\mathbb Z_2$ symmetry explicitly
on the MPS tensors, we target states of definite parity by introducing a
parity bias in the Hamiltonian.

To obtain the lowest odd state, we minimize
\begin{equation}
	H_{\rm odd}
	=
	H_{\mathrm{lat}}+w_P\hat P,
	\qquad
	w_P>0 .
	\label{eq:odd_target_hamiltonian}
\end{equation}
Since odd states have parity $P=-1$, their energies are shifted by
$-w_P$, while even states are shifted by $+w_P$. Choosing $w_P$
sufficiently large therefore makes the lowest odd state the ground state
of $H_{\rm odd}$.  Since $[H_{\rm lat},\hat P]=0$, the eigenstates are unaffected. Once the corresponding MPS
$|\Psi_{\rm odd}\rangle$ has been obtained, its physical energy is
computed as
\begin{equation}
	E_{\rm odd}
	=
	\frac{
		\langle\Psi_{\rm odd}|H_{\mathrm{lat}}|\Psi_{\rm odd}\rangle
	}{
		\langle\Psi_{\rm odd}|\Psi_{\rm odd}\rangle
	}.
	\label{eq:odd_physical_energy}
\end{equation}

The lowest non-trivial even state is obtained similarly. We use
\begin{equation}
	H_{\rm even}
	=
	H_{\mathrm{lat}}-w_P\hat P,
	\label{eq:even_target_hamiltonian}
\end{equation}
which lowers the even sector relative to the odd sector. Since the vacuum
itself is even, the variational state must additionally be orthogonal to
the previously obtained ground state,
\begin{equation}
	\langle\Psi_0|\Psi_{\rm even}\rangle=0 .
	\label{eq:even_orthogonality}
\end{equation}
The energy of the state $|\Psi_{\rm even}\rangle$ is then computed as
\begin{equation}
	E_{\rm even}
	=
	\frac{
		\langle\Psi_{\rm even}|H_{\mathrm{lat}}|\Psi_{\rm even}\rangle
	}{
		\langle\Psi_{\rm even}|\Psi_{\rm even}\rangle
	}.
	\label{eq:even_physical_energy}
\end{equation}

An alternative would be to implement the $\mathbb Z_2$ symmetry directly
in the MPS tensors and optimize separately within fixed parity sectors.
We do not use a symmetry-resolved MPS representation here. Instead, all
optimizations are carried out in the full MPS manifold using the
parity-biased Hamiltonians above. We perform these optimizations using the density-matrix
renormalization-group (DMRG) algorithm
\cite{white1992DensityMatrix,schollwoeck2011DensitymatrixRenormalization}.

\subsection{Entanglement in AdS and MPS efficiency}
\label{sec:ads_mps_entanglement}

Let us now understand how the entanglement behaves for the two kinds of
problems considered in this work: extracting the low-lying spectrum of a
massive theory and studying bulk criticality. As discussed above, the
efficiency of an MPS depends on how the entanglement grows as the relevant
continuum or scaling limit is approached.

For a relativistic QFT in one spatial dimension, when the UV length
scale is much smaller than the physical correlation length,
$a\ll\xi$, the entanglement across a spatial cut has the familiar
logarithmic dependence \cite{calabrese2004EntanglementEntropy},
\begin{equation}
	S_{\mathrm{cut}}
	=
	\alpha
	\log\left(\frac{\xi}{a}\right)
	+\mathrm{const.}
	+\cdots ,
	\label{eq:entropy_correlation_length}
\end{equation}
For a gapped local one-dimensional Hamiltonian, $\xi$ remains
finite, so the entropy approaches a constant as the physical size of the
system is increased at fixed $a$. This is the one-dimensional
entanglement area law \cite{hastings2007AreaLaw}. The low-lying excited
states considered have the same leading UV
entanglement structure as the ground state, differing only by finite state-dependent contributions. Their entanglement therefore has the same parametric scaling. 

In global AdS, the corresponding length scale can be read directly from
the physical energy gap,
\begin{equation}
	\xi
	\sim
	\frac{1}{E_{\mathrm{gap}}}
	=
	\frac{R}{\Delta_{\mathrm{gap}}}.
	\label{eq:ads_correlation_length}
\end{equation}
Which gap appears depends on the observable being considered. For
example, correlations of the field are controlled by the lowest
$\mathbb Z_2$-odd gap, while even observables couple to the even spectrum.

The same scale can also be seen directly from the proper-distance
falloff. At large $|\rho|$,
\begin{equation}
	\cos r
	=
	\operatorname{sech}\left(\frac{\rho}{R}\right)
	\sim
	2e^{-|\rho|/R}.
\end{equation}
A contribution of dimension $\Delta_{\mathrm{gap}}$ in the
bulk--boundary OPE therefore falls off as
\begin{equation}
	(\cos r)^{\Delta_{\mathrm{gap}}}
	\sim
	e^{-\Delta_{\mathrm{gap}}|\rho|/R},
\end{equation}
again giving a characteristic length scale
$R/\Delta_{\mathrm{gap}}$.

\paragraph{\textbf{Massive theories in the continuum limit:}}
For a massive theory at large $R$, the dimensionless gap grows as
\begin{equation}
	\Delta_{\mathrm{gap}}\sim m_{\rm phys}R,
\end{equation}
and hence
\begin{equation}
	\xi \sim
	\frac{R}{\Delta_{\mathrm{gap}}}
	\sim
	\frac{1}{\rm m_{phys}}.
\end{equation}
The physical correlation length therefore approaches a constant as
$R$ is increased. Once $R\gg\xi$, increasing the AdS radius or moving
the radial cutoff farther away does not appreciably affect the
entanglement across a cut near the centre.

Now let us consider the continuum limit $a/R << 1$ for fixed $R$. In uniform global-coordinate discretization, the proper lattice spacing
near the centre is $a\simeq R a_r$. Equation~\eqref{eq:entropy_correlation_length}
then gives
\begin{equation}
	S_{\mathrm{mid}}
	=
	\alpha
	\log\left(\frac{1}{a_r}\right)
	+\mathrm{const.}
	+\cdots ,
	\label{eq:massive_entanglement_scaling}
\end{equation}
where the dependence on $\Delta_{\mathrm{gap}}$ has been absorbed into the constant. Therefore, as we take the continuum limit, $a_r \to 0$,  the entanglement grows logarithmically. The same leading logarithmic scaling applies to the low-lying excited states used to extract the boundary spectrum.

\paragraph{\textbf{Bulk criticality:}}
At a critical point, as $R$ is increased, the dimensionless gap approaches a finite universal value,
\begin{equation}
	\Delta_{\mathrm{gap}}
	\longrightarrow
	\Delta_{\mathrm{gap}}^{*}.
\end{equation}
Equation~\eqref{eq:ads_correlation_length} then gives
\begin{equation}
	\xi
	\sim
	\frac{R}{\Delta_{\mathrm{gap}}^{*}}
	\propto R,
\end{equation}
so the AdS radius itself sets the correlation length at criticality.

For the proper-distance lattice used in the finite-size-scaling
analysis, the lattice spacing is simply $a=a_\rho$.  In the scaling limit $R/a_{\rho} >>1$ and $a_{\rho}$ fixed,
Eq.~\eqref{eq:entropy_correlation_length} gives
\begin{equation}
	S_{\mathrm{mid}}
	=
	\alpha
	\log\left(
	R
	\right)
	+\mathrm{const.}
	+\cdots .
	\label{eq:critical_ads_entanglement}
\end{equation}
where  the dependence on $\Delta_{\mathrm{gap}}^{*}$ and $a_\rho$ has been absorbed into the constant. The entanglement therefore grows only logarithmically with $R$. The same leading scaling applies to the low-lying
excited states at criticality.

In both cases, we find that the entanglement grows at most logarithmically in the
limit being taken. The bond dimension can therefore be increased
systematically as the continuum or critical limit is approached,
while remaining far smaller than the dimension of the full
tensor-product Hilbert space. This is what makes the MPS description
practical for both the boundary-spectrum and bulk-scaling calculations considered in this work.

% -----------------------------------------------------------------------------
\section{Scalar $\phi^4$ theory in AdS$_2$}
\label{sec:phi4}
% -----------------------------------------------------------------------------
We now apply the general construction of the previous sections to
$\mathbb Z_2$-invariant scalar $\phi^4$ theory in $\AdS_2$. We first specify the renormalized Hamiltonian and the boundary conditions. We then study the low-lying global-AdS spectrum, where the continuum energy gaps determine  the boundary scaling dimensions. Bulk criticality and the corresponding conformal boundary condition will be studied separately in Sec.~\ref{sec:criticality}.

\subsection{Hamiltonian and mass renormalization}
\label{sec:phi4_renormalization}
In two spacetime dimensions, $\phi^4$ theory is superrenormalizable and,
apart from an additive vacuum-energy renormalization, requires only a
mass renormalization
\cite{rychkov2015HamiltonianTruncation,rychkov2016HamiltonianTruncation}. We can write the renormalized interaction density  in \eqref{eq:continuum_ads_hamiltonian} as
\begin{equation}
  \mathcal U(r,\phi)
  =
  \frac12
  \left[
    m^2+\delta m^2
  \right]\phi^2
  +
  \lambda\phi^4 ,
  \label{eq:phi4_interaction_with_counterterm}
\end{equation}
where $m^2$ and $\lambda$ denote the renormalized dimensionful
couplings and  $\delta m^2$ is the mass counterterm that we fix below. We could also include the curvature coupling term $\kappa \mathcal{R}\phi^2$, but since the scalar curvature of $\mathrm{AdS}_2$ is a constant, $\mathcal{R}=-2/R^2$, it can always be absorbed into the definition of the mass parameter. We therefore set $\kappa =0$.

\paragraph{\textbf{Mass Renormalization:}} 
A natural prescription is to normal-order the interaction with respect to the free AdS vacuum. Let $G_{\rm AdS}(x,x';m_{\rm ref}^2)$
denote the free scalar two-point function on AdS$_2$ with radius $R$ and a reference mass $m_{\rm ref}^2$. The corresponding
AdS tadpole is formally defined as the coincident limit
\begin{equation}
  Z_{\rm AdS}(m_{\rm ref}^2)
  \equiv
  \lim_{x'\rightarrow x}
  G_{\rm AdS}(x,x';m_{\rm ref}^2)\,.
\end{equation}
Normal-ordering with respect to the AdS vacuum then corresponds to the subtraction
\begin{equation}
  \delta m^2_{AdS}
  =
  -12\lambda
  Z_{\rm AdS}(m_{\rm ref}^2) \,,
\end{equation}
where the factor of $12$ follows from the normalization
$\lambda\phi^4$. This freedom of choosing an arbitrary reference mass is possible because the UV divergence is independent of the scalar
mass. On the
lattice, the regulated AdS tadpole becomes position dependent. We denote it by  $Z^{(lat)}_{\rm AdS}(r_j; m_{\rm ref}^2)$.

The prescription above is perfectly valid at a fixed radius. In curved
space, however, the finite part of the renormalization prescription may
contain curvature-dependent contributions
\cite{banados2023BulkRenormalization}. The difficulty for our purposes arises when we vary the AdS radius. The AdS tadpole contains an $R$-dependent finite contribution, so normal-ordering separately at each $R$ changes
the relation between the bare and renormalized mass. As a result,
holding $m^2$ and $\lambda$ fixed while varying $R$ would not correspond to placing the same renormalized QFT on
different AdS backgrounds. This is precisely what we need for the
finite-size-scaling analysis of Sec.~\ref{sec:criticality}.

We therefore fix the finite part of the mass subtraction by matching to a
flat-space reference defined with the same lattice regulator. Let
$Z_{\rm flat}^{(\mathrm{lat})}(m_{\rm ref}^2)$ denote the corresponding
flat-space tadpole, computed with the same reference mass and UV
resolution as the AdS lattice. At the centre of AdS we then define
\begin{equation}
	z_R^{(\mathrm{lat})}(m_{\rm ref}^2)
	=
	Z_{\rm AdS}^{(\mathrm{lat})}(0;m_{\rm ref}^2)
	-
	Z_{\rm flat}^{(\mathrm{lat})}(m_{\rm ref}^2).
	\label{eq:zR_definition}
\end{equation}
This finite difference fixes the curvature-dependent finite part by matching the AdS theory to the corresponding flat-space
renormalisation scheme.

For the global-coordinate lattice, we define
\begin{equation}
	z_R(m_{\rm ref}^2)
	=
	\lim_{\substack{a_r\to0\\ r_{\max}\to\pi/2}}
	z_R^{(\mathrm{lat})}(m_{\rm ref}^2),
\label{eq:flat_corr_global}
\end{equation}
while for the proper-distance lattice used in the finite-size-scaling
analysis,
\begin{equation}
	z_R(m_{\rm ref}^2)
	=
	\lim_{\substack{a_\rho\ \mathrm{fixed}\\ \rho_{\max}\to\infty}}
	z_R^{(\mathrm{lat})}(m_{\rm ref}^2).
\label{eq:flat_corr_proper}
\end{equation}
 From here onwards, $z_R$ should be understood as the regulator-specific limit appropriate to the discretization being used.
Our renormalization prescription is then
\begin{equation}
	\delta m^2(r_j)
	=
	12\lambda
	\left[
	-Z_{\rm AdS}^{(\mathrm{lat})}(r_j;m_{\rm ref}^2)
	+z_R(m_{\rm ref}^2)
	\right].
	\label{eq:mass_counterterm_scheme}
\end{equation}
The first term removes the local AdS tadpole, while $z_R$ fixes the
finite part by matching to the corresponding flat-space regulator.
For the proper-distance discretization used in the finite-size-scaling
analysis, the fixed spacing $a_\rho$ ensures that the same flat-space
lattice renormalization scheme is used for every $R$.

For all numerical results in this paper, we choose 
\begin{equation}
      m_{\rm ref}^2=m^2\,.
\end{equation}
The full renormalized interaction density of the $\phi^4$ theory  on the lattice becomes
\begin{equation}
	\mathcal{U}(r_j,\phi_j)
	=
	\frac{1}{2}
	\left[
	m^2
	+
	12\lambda
	\left(
	-Z_{\rm AdS}^{(\mathrm{lat})}(r_j;m^2)
	+z_R(m^2)
	\right)
	\right]\phi_j^2
	+
	\lambda\phi_j^4 .
	\label{eq:interaction_density_lattice}
\end{equation}

To obtain the corresponding renormalized Hamiltonian, we simply plug the above expression in Eq.~\eqref{eq:lattice_hamiltonian}. 

\paragraph{\textbf{Boundary conditions:}}
Throughout this work, we impose Dirichlet boundary conditions at the two asymptotic boundaries of global AdS,
\begin{equation}
    \phi\!\left(r =\pm\frac{\pi}{2}\right)=0.
\end{equation}
On the regulated lattice, this is implemented by fixing the endpoint fields to zero,
\begin{equation}
    \widetilde{\phi}_1=\widetilde{\phi}_N=0,
\end{equation}
so that only the interior sites are treated as dynamical degrees of freedom.

\paragraph{\textbf{Continuum and lattice tadpole computation:}} The discussion above specifies the renormalisation prescription but not the explicit evaluation of the AdS and flat-space tadpoles. In the continuum, these are obtained from the normal-mode expansion of the free scalar field by constructing the equal-time two-point function and taking its regulated coincident
limit. On the lattice, the corresponding quantities are computed from the quadratic lattice Hamiltonian using its free normal modes.  Details of the free-field mode expansion and the corresponding  lattice tadpole calculations are given in
Appendix~\ref{app:renormalization}.

\subsection{Extracting the boundary spectrum}
\label{sec:boundary_spectrum}

 In this section, we focus on the lowest non-trivial boundary primaries in the
$\mathbb Z_2$-odd and $\mathbb Z_2$-even sectors, obtained from the
corresponding low-lying eigenstates of the renormalised Hamiltonian of $\phi^4$ theory. The free theory provides a natural starting point for identifying these
levels. For a scalar field in AdS$_2$, the lowest $\mathbb Z_2$-odd
boundary primary has scaling dimension
\cite{carmi2019StudyQuantum}
\begin{equation}
    \Delta^{(free)}_{\rm odd}=\Delta_\phi,
    \qquad
    \Delta_\phi(\Delta_\phi-1)= m^2 R^2,
\end{equation}
while the lowest non-trivial $\mathbb Z_2$-even boundary primary
associated with the bulk operator $\phi^2$ has scaling dimension
\begin{equation}
    \Delta^{(free)}_{\rm even}=2\Delta_\phi.
\end{equation}

Turning on the quartic coupling deforms the free theory and continuously
changes the boundary spectrum. Since the interaction preserves the
$\mathbb Z_2$ symmetry, the odd and even sectors remain distinct, and
the lowest levels in each sector can be followed continuously from their
free-theory counterparts associated with $\phi$ and $\phi^2$ into the
interacting regime. The calculations below therefore provide a direct
non-perturbative determination of how the low-lying boundary spectrum
evolves with the interaction strength.

The continuum spectrum depends only on
the dimensionless parameters $\bar m^2= m^2 R^2$ and $\bar\lambda = \lambda R^2$. Throughout this section we fix
    $\bar m^2 = 1$
and study the spectrum as a function of the dimensionless quartic
coupling $\bar\lambda$. In the numerical implementation we set $R=1$, which simply fixes the overall unit of length.

We use the uniform global-coordinate discretisation introduced in
Eq.~\eqref{eq:uniform_global_grid}.
In all the results of this section that follow, we choose the radial cutoff $\epsilon$ small enough and likewise, the local Hilbert-space dimension $d$ and bond dimension $\chi$ large enough so that their errors are smaller than the finite-$N$ discretisation errors, and demonstrate the convergence of our results with $N$. The numerical setup is summarised in Table~\ref{tab:boundary_spectrum_setup}.

Let us point out that, already at $\bar\lambda=0$, reproducing the known free spectrum provides
a non-trivial check of the lattice and MPS calculation. At weak couplings, we further compare the numerical spectrum with perturbation theory
through $O(\bar\lambda^2)$, before following the same boundary levels into the non-perturbative regime.

\begin{table}[h!]
\centering
\caption{Numerical setup used for the boundary-spectrum calculations.}
\label{tab:boundary_spectrum_setup}
\begin{tabular}{ll}
\toprule
Renormalized mass & $\bar m^2 = 1$ \\
Spatial discretisation &
Global-coordinate lattice, Eq.~\eqref{eq:uniform_global_grid} \\
Radial cutoff &
$\epsilon=10^{-3}$ \\
Local basis &
Free-vacuum basis, Eq.~\eqref{eq:free_vacuum_frequency} \\
Local Hilbert-space dimension &
$d = 16$ \\
Maximum bond dimension &
$\chi= 80$ \\
\bottomrule
\end{tabular}
\end{table}

\subsubsection{Weak-coupling results and perturbative checks}
At weak coupling, the boundary scaling dimensions admit a perturbative expansion in $\bar \lambda$,
\begin{equation}
    \Delta_i(\bar\lambda)
    =
    \Delta_i^{(0)}
    +
    \bar\lambda\,\Delta_i^{(1)}
    +
    \bar\lambda^2\,\Delta_i^{(2)}
    + O(\bar\lambda^3).
\end{equation}
For the lowest odd and even levels, we obtain the perturbative coefficients upto
$\mathcal O(\bar\lambda^2)$ from standard
Rayleigh--Schr\"odinger perturbation theory applied to the global AdS Hamiltonian discretized on the global-coordinate lattice. Details are provided in Appendix~\ref{app:perturbation_theory}. Related perturbative calculations of global-AdS energy
levels were studied systematically in the Hamiltonian-truncation
framework of Ref.~\cite{hogervorst2021HamiltonianTruncation}. 

 \begin{figure}[h!]
	\centering
	\includegraphics[width=1\linewidth]{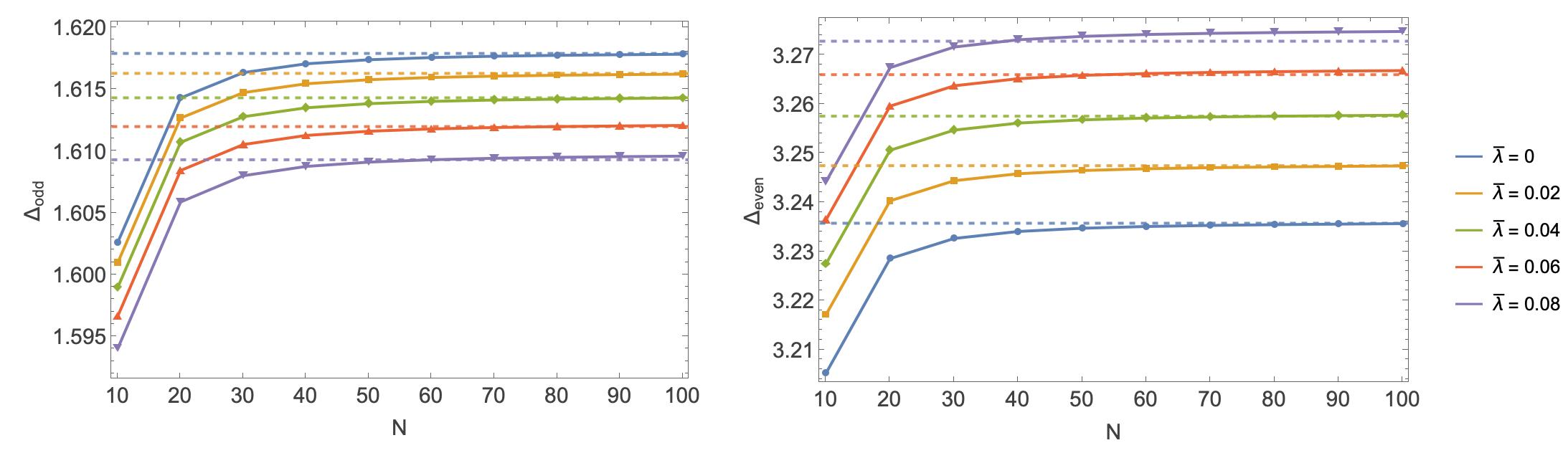}
	\caption{Convergence in the continuum limit of the lowest
		$\mathbb Z_2$-odd (left) and $\mathbb Z_2$-even (right) boundary
		scaling dimensions at weak coupling, for $\bar m^2=1$. Dashed lines
		denote the $O(\bar\lambda^2)$ perturbative predictions.}
	\label{fig:ContMassGapOddEvenWeak}
\end{figure}

Figure~\ref{fig:ContMassGapOddEvenWeak} compares the boundary scaling dimensions extracted from the MPS calculation with the perturbative predictions. As the lattice is refined, the numerical results approach the perturbative values in both symmetry sectors. The agreement is already
good at moderate $N$, while the small residual differences at the
largest couplings shown are consistent with corrections beyond the
$O(\bar\lambda^2)$ truncation. Together with the exact
$\bar\lambda=0$ spectrum, this provides a simple check that the
discretised Hamiltonian, renormalisation prescription, and MPS
calculation reproduce the expected weak-coupling physics.

\subsubsection{Beyond weak coupling}
\begin{figure}[h!]
    \centering
    \includegraphics[width=1\linewidth]{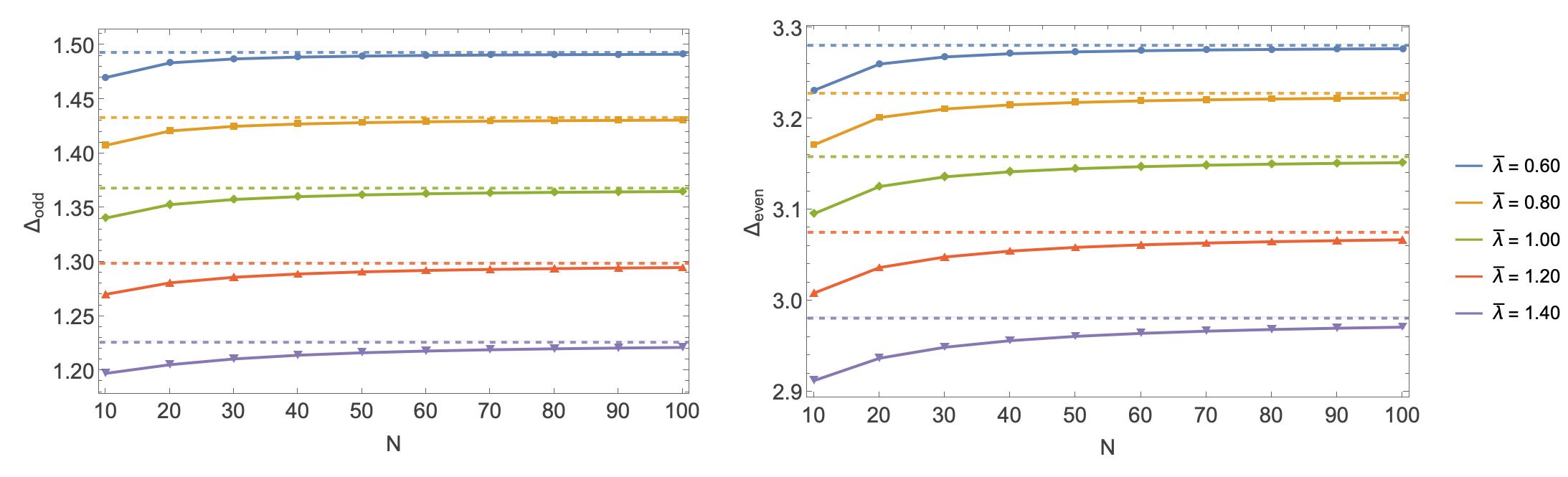}
    \caption{Convergence in the continuum limit of the lowest $\mathbb Z_2$-odd (left) and $\mathbb Z_2$-even (right) boundary scaling dimensions beyond the weak-coupling regime, for $\bar m^2=1$. Dashed lines are continuum estimates from the fit $\Delta(N)=\Delta_{\rm cont} + a/N^2.$
    }
    \label{fig:ContMassGapOddEvenStrong}
\end{figure}

We now follow the same boundary levels beyond the regime where
low-order perturbation theory is quantitatively reliable.
Figure~\ref{fig:ContMassGapOddEvenStrong} shows the approach to the
continuum for several representative couplings. As the couplings grow larger, convergence in the continuum limit becomes slower, but the
finite-$N$ dependence remains smooth and the continuum values can still
be extracted reliably. For the continuum extrapolation, we use
\begin{equation}
	\Delta(N)
	=
	\Delta_{\rm cont}
	+\frac{a}{N^2}.
\end{equation}
At fixed radial cutoff, the lattice spacing scales as $a_r\sim 1/N$, and
the leading discretisation errors are therefore expected to be $O(a_r^2)\sim O(1/N^2)$ at large $N$. We estimate the uncertainty in $\Delta_{\rm cont}$
by varying the range of lattice sizes included in the fit; the resulting error bars are smaller than the plotting resolution in Fig.~\ref{fig:ContMassGapOddEvenStrong}.

\begin{figure}[h!]
	\centering
	\includegraphics[width=1\linewidth]{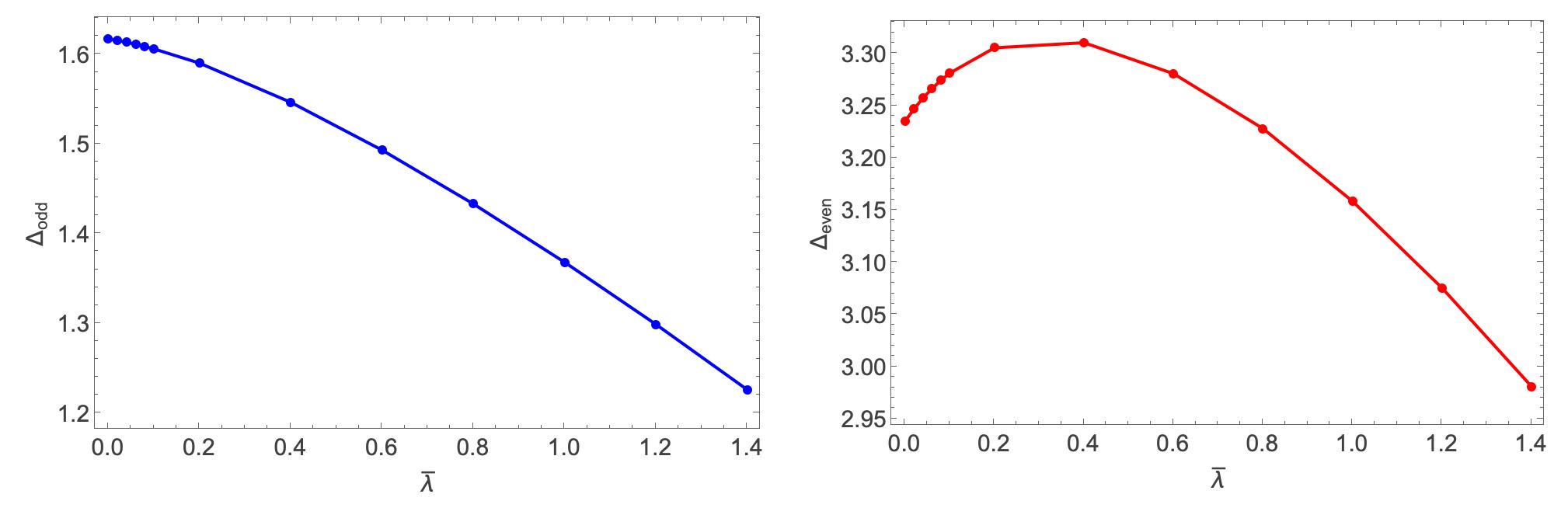}
	\caption{Lowest $\mathbb Z_2$-odd (left) and $\mathbb Z_2$-even (right) boundary scaling dimensions as functions of the quartic coupling $\bar{\lambda}$ for $\bar m^2=1$.}
	\label{fig:ContMassGapOddEvenFinal}
\end{figure}

The final continuum-extrapolated boundary scaling dimensions as functions of the quartic coupling $\bar \lambda$ are shown in Fig.~\ref{fig:ContMassGapOddEvenFinal}.  The lowest $\mathbb Z_2$-odd dimension decreases smoothly with
$\bar\lambda$. The lowest $\mathbb Z_2$-even level starts from its free-theory value
$2\Delta_\phi$, initially increases, and then turns over and begins to
decrease. This is qualitatively consistent with second-order
perturbation theory, where the $O(\bar\lambda^2)$ correction is
negative, although perturbation theory predicts the turnover at a smaller coupling and a much faster subsequent decrease.

% -----------------------------------------------------------------------------
% -----------------------------------------------------------------------------
\section{Bulk criticality and conformal boundary conditions}
\label{sec:criticality}
% -----------------------------------------------------------------------------

We now turn to the bulk $\mathbb Z_2$ symmetry-breaking transition of
the interacting $\phi^4$ theory. As discussed in
Sec.~\ref{sec:bulk_boundary_criticality}, the AdS radius acts as an
infrared scale, so varying $R$ at fixed renormalised dimensionful
couplings allows us to perform finite-size scaling directly in AdS.
There are two questions we would like to answer. First, can we locate the bulk transition expected to lie in the 2D Ising universality class and extract the corresponding universal critical data? Second, which conformal boundary condition is reached at the same critical point?

For $\mathbb Z_2$-invariant $\phi^4$ theory in two
dimensions,  the continuous symmetry-breaking transition is governed by the critical
Ising CFT. This flow has been studied non-perturbatively using
Hamiltonian truncation, tensor-network methods, and related numerical
approaches
\cite{rychkov2015HamiltonianTruncation,anand2017RGFlow,
	milsted2013MatrixProduct,kadoh2019TensorNetwork}.
The two relevant bulk operators are the $\mathbb Z_2$-odd spin field
and the $\mathbb Z_2$-even energy operator, with
\begin{equation}
    \Delta_\sigma=\frac18,
    \qquad
    \Delta_\epsilon=1 .
\end{equation}
Correspondingly, the familiar Ising critical exponents include
$\nu=1$ and $\beta=1/8$. These will provide the basic targets for the
finite-size-scaling analysis below.

At the critical point there is also universal boundary data. As
reviewed earlier, a CFT on AdS$_2$ is related by a Weyl transformation
to a BCFT on a flat strip, with the two asymptotic AdS boundaries
mapped to the two edges of the strip. The global-AdS energy gaps above the vacuum can therefore be interpreted as the spectrum of boundary operators and give direct access to the conformal boundary condition. This point of view was also used in the Hamiltonian-truncation study of
Ref.~\cite{hogervorst2021HamiltonianTruncation}.

The 2D Ising CFT admits three elementary conformal
boundary conditions: free, fixed $+$, and fixed $-$
\cite{cardy1984ConformalInvariance,cardy1986EffectBoundary,
	cardy1989BoundaryConditions}. The two fixed boundary conditions select a
sign for the order parameter and are exchanged by the global
$\mathbb Z_2$ symmetry. The free boundary condition, on the other hand,
preserves the $\mathbb Z_2$ symmetry. In the language of boundary
critical phenomena, this symmetry-preserving fixed point is usually
referred to as the \emph{ordinary} Ising boundary condition. We will
therefore use \emph{free/ordinary} for this boundary condition below,
and simply refer to the other two as the fixed or symmetry-breaking
boundary conditions.

Our microscopic Dirichlet condition,
$\phi\big|_{\partial {\rm AdS}}=0$, also preserves the
$\mathbb Z_2$ symmetry. At the bulk critical point, we therefore expect
it to flow to the symmetry-preserving free/ordinary Ising conformal
boundary condition. For this boundary condition, the lowest non-trivial
$\mathbb Z_2$-odd and $\mathbb Z_2$-even global boundary primaries have
scaling dimensions
\cite{cardy1986EffectBoundary,cardy1989BoundaryConditions}
\begin{equation}
	\Delta_{\rm odd}^{(\mathrm{Ising})}
	=
	\frac12,
	\qquad
	\Delta_{\rm even}^{(\mathrm{Ising})}
	=
	2.
	\label{eq:ising_boundary}
\end{equation}

These dimensions refer to primaries of the global $SL(2,\mathbb R)$
conformal algebra. In the full Ising BCFT, the odd state is also a
Virasoro primary, whereas the even state belongs to the identity
Virasoro family and is a Virasoro descendant
\cite{cardy1986EffectBoundary,cardy1989BoundaryConditions}. By first locating
the bulk critical point and then comparing the low-lying boundary
spectrum there with Eq.~\eqref{eq:ising_boundary}, we can directly test
whether the Dirichlet boundary condition flows to the expected
free/ordinary Ising boundary condition. 

\subsection{Bulk criticality in $\phi^4$ theory}
\label{sec:binder}
% -----------------------------------------------------------------------------

Before explaining how we probe bulk criticality in $\phi^4$ theory numerically, it is useful to recall the nature of spontaneous symmetry breaking in AdS.
A constant-time slice of global AdS has infinite proper volume. Spontaneous symmetry breaking is therefore not excluded even at finite $R$, despite the discrete energy spectrum. As discussed in
Ref.~\cite{hogervorst2021HamiltonianTruncation}, symmetry breaking in AdS has an
``all-or-nothing'' character. Given symmetry-preserving boundary conditions, a configuration in which the broken
phase occupies only a finite region around the centre and returns to the symmetric phase sufficiently close to the boundary, is
not a stable endpoint. If the broken phase is energetically preferred, the ordered region instead extends arbitrarily far towards the boundary, and only goes to the symmetric phase asymptotically. The resulting broken vacua form distinct superselection sectors, much as in
infinite-volume flat space.

\subsubsection{Central bulk observables}
\label{sec:central_bulk_observables}
On the lattice in AdS$_2$, the above picture becomes important since the infinite spatial slice must be truncated at a finite proper distance, $  |\rho|\leq \rho_{\max}$,
and the ($\mathbb{Z}_2$ symmetry-preserving) Dirichlet condition is then imposed at these regulated
endpoints. Unlike the AdS radius $R$, the radial cutoff $\rho_{\max}$ is not a physical IR regulator and is only introduced by our numerical
discretization. If the radial cut-off is not taken to be sufficiently far away, it can influence observables in the
interior by suppressing ordering. At fixed $R$, we must therefore consider a sufficiently large
$\rho_{\max}$ so that  observables measured close to the centre are insensitive to it. We therefore restrict the measurements entering the bulk analysis to a
central window
\begin{equation}
	\mathcal W_R=\{\,\rho:\ |\rho|<\rho_{\rm bulk}\,\},
	\label{eq:bulk_window}
\end{equation}
with $\rho_{\rm bulk}$ chosen to scale with $R$. The window then grows
with the AdS radius, as required for finite-size scaling, while remaining
well separated from the regulated endpoints. This allows us to probe an
increasingly large bulk region without including the part of the lattice
most strongly affected by the radial cutoff.

The natural order parameter in our case is the spatial average of the scalar field over this central region,
\begin{equation}
    M
    \equiv
    \frac{1}{2 \rho_{bulk}}
    \int_{-\rho_{bulk}}^{\rho_{bulk}} d\rho\,\phi(\rho).
    \label{eq:bulk_order_parameter}
\end{equation}

On a finite regulated lattice, however, a $\mathbb Z_2$-symmetric
ground state satisfies
\begin{equation}
	\langle M\rangle=0
\end{equation}
in both the disordered and ordered regimes.
True spontaneous symmetry
breaking only appears in the full AdS theory when the radial cutoff is removed. We therefore use the fluctuations of $M$ to detect ordering. We work with the even moments
\begin{equation}
    \langle M^2\rangle,
    \qquad
    \langle M^4\rangle \,.
\end{equation}
A particularly useful dimensionless combination of these moments is
the Binder cumulant \cite{binder1981CriticalProperties},
\begin{equation}
    U
    =
    1-
    \frac{\langle M^4\rangle}
         {3\langle M^2\rangle^2}.
    \label{eq:binder_definition}
\end{equation}
The overall normalization of $M$ cancels in this ratio, and $U$ therefore provides a dimensionless probe of the shape of the order-parameter distribution. This makes it especially useful for finite-size scaling as we shall see in the next section. Deep in the symmetric phase, $M$ fluctuates around zero and its distribution becomes approximately Gaussian. In this
limit, $
    \langle M^4\rangle
    \simeq
    3\langle M^2\rangle^2$, 
and hence 
\begin{equation}
U\simeq0
\end{equation}
On the ordered side, the distribution instead develops two peaks associated with the two symmetry-related bulk vacua. Deep in the ordered phase these peaks become narrow around $M=\pm M_0$,  and therefore
\begin{equation}
    U\simeq\frac23. 
\end{equation}
Thus the Binder cumulant provides a particularly transparent diagnostic of the transition, interpolating between the characteristic values $0$ and $2/3$ in the symmetric and ordered regimes. These
should be understood as the limiting values in the two phases, rather than as strict bounds on the Binder cumulant.

\subsubsection{Extracting critical data from finite-size scaling}
\label{sec:ads_fss}

As discussed in Sec.~\ref{sec:bulk_boundary_criticality}, we can study
the phase transition of the flat-space $\phi^4$ theory by placing it on AdS backgrounds of increasing radii and
analysing the finite-size scaling of the measured quantities in $R$. The renormalized dimensionful
couplings $m^2$ and $\lambda$ must therefore be held fixed as $R$ is
varied. This is also where the mass-renormalization prescription
introduced in Sec.~\ref{sec:phi4_renormalization} becomes important. The
finite part of the mass subtraction is fixed by matching to the
corresponding flat-space prescription. This removes an otherwise
$R$-dependent finite curvature contribution and ensures that fixed
values of the renormalized $m^2$ and $\lambda$ refer to the same QFT at
different AdS radii. Varying $R$ can therefore be interpreted as changing
the background geometry rather than the microscopic theory. In the flat-space limit at large $R$,  the physics is controlled by the dimensionless ratio
$\lambda/m^2$.  To locate the critical point, we therefore fix $m^2$ and scan $\lambda$ through the transition.

Close to a continuous transition, the bulk correlation length and
finite-size dependence are governed by the usual scaling hypothesis
\cite{fisher1972ScalingTheory,campostrini2014FinitesizeScaling}.
The correlation length diverges as
\begin{equation}
	\xi\sim |t|^{-\nu},
	\qquad
	t\equiv\frac{\lambda-\lambda_c}{\lambda_c},
\end{equation}
where the normalization of the reduced coupling $t$ is conventional.
Since $R$ provides the IR scale, the corresponding scaling
variable is
\begin{equation}
	x
	=
	t\left(\frac{R}{a_\rho}\right)^{1/\nu}.
	\label{eq:fss_variable}
\end{equation}
The lattice spacing $a_\rho$ is held fixed throughout the scaling
analysis.

For a dimensionless quantity such as the Binder cumulant, the leading
finite-size dependence is entirely through $x$ \cite{binder1981CriticalProperties,campostrini2014FinitesizeScaling},
\begin{equation}
	U(\lambda,R)
	=
	F_U(x)
	+
	\left(\frac{R}{a_\rho}\right)^{-\omega}G_U(x)
	+\cdots ,
	\label{eq:binder_fss}
\end{equation}
where the second term denotes the leading correction to scaling.
Exactly at the critical point, $x=0$, and the leading value of $U$ is
therefore independent of $R$. Binder curves at increasing radii should
thus cross progressively closer to the critical coupling.

We now implement this scaling analysis numerically. We fix
$m^2=0.01$ and $a_\rho=1$, vary the quartic coupling through the
transition, and repeat the calculation for a sequence of increasing
AdS radii. For each $R$, we take $\rho_{\max}=5R$ and do measurements in the central region $|\rho|<R$. The remaining
numerical parameters are chosen such that their residual effects are
smaller than the finite-$R$ dependence relevant for the scaling
analysis. The setup is summarized in
Table~\ref{tab:bulk_criticality_setup}.

\begin{table}[h!]
	\centering
	\caption{Numerical setup used for the bulk finite-size-scaling analysis.}
	\label{tab:bulk_criticality_setup}
	\begin{tabular}{ll}
		\toprule
		Renormalized mass &
		$m^2=0.01$ \\
		Spatial discretisation &
		Proper-distance lattice, Eq.~\ref{eq:proper_distance_grid} \\
		Lattice spacing &
		$a_\rho=1$ \\
		Radial cutoff &
		$\rho_{\max}=5R$ \\
		Bulk measurement window &
		$\rho_{\rm bulk}=R$ \\
		Local basis &
		On-site quadratic basis, Eq.~\ref{eq:onsitebasis} \\
		Local Hilbert-space dimension &
		$d=10$ \\
		Maximum bond dimension &
		$\chi=50$ \\
		\bottomrule
	\end{tabular}
\end{table}

\begin{figure}[b!]
	\centering
	\includegraphics[width=0.75\linewidth]{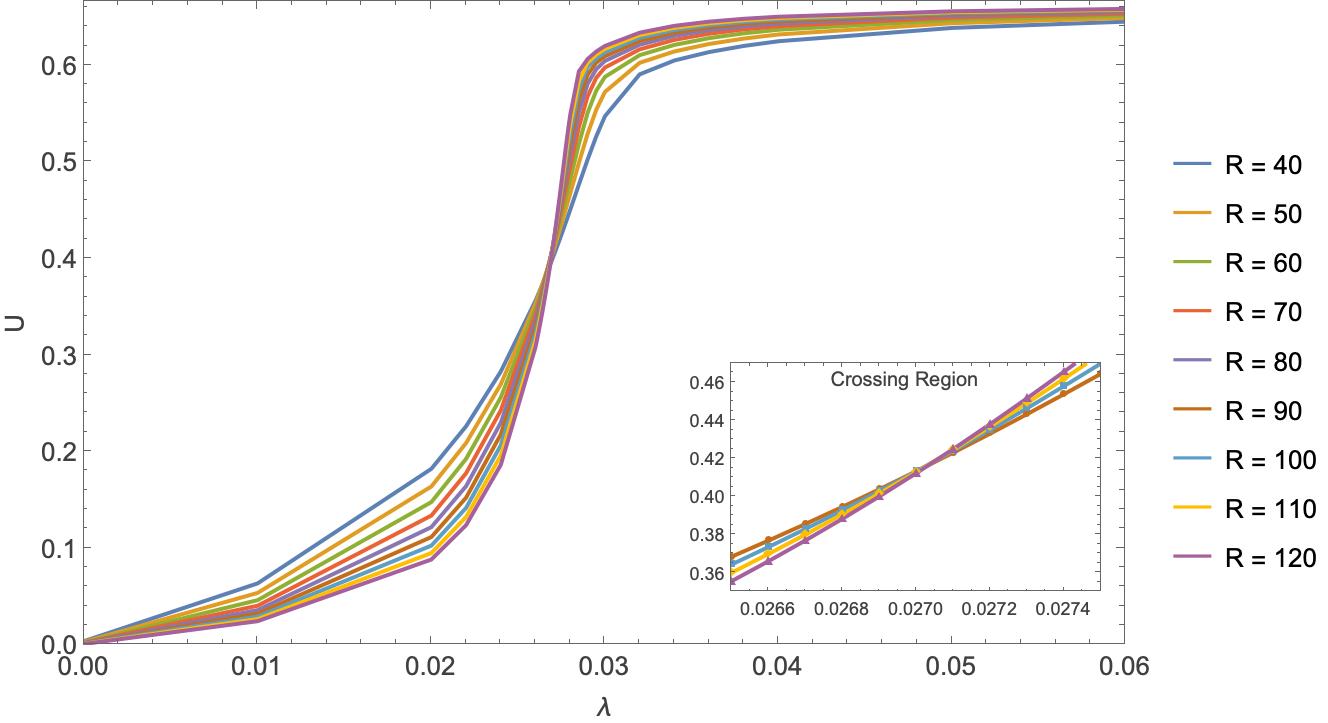}
	\caption{
		Binder cumulant $U$ of the central bulk order parameter as a
		function of the quartic coupling $\lambda$ for $m^2=0.01$ and
		increasing AdS radius $R$. The order parameter is measured within
		$|\rho|<R$, while the regulated boundaries are placed at
		$\rho_{\max}=5R$. The evolution from $U\simeq0$ in the symmetric
		regime towards $U\simeq2/3$ in the ordered regime becomes
		increasingly sharp with $R$. The inset magnifies the crossing
		region for the largest radii.
	}
	\label{fig:binder_crossings}
\end{figure}

With this setup, Fig.~\ref{fig:binder_crossings} shows the Binder
cumulant as a function of $\lambda$ for several AdS radii. We see that $U$ evolves from $U\simeq0$ in the symmetric regime to
$U\simeq2/3$ in the ordered regime. As $R$ increases, this crossover
becomes progressively sharper, narrowing the finite-$R$ transition region. Moreover, the curves for larger radii also develop stable crossings.  The inset of Fig.~\ref{fig:binder_crossings} shows this crossing region
more closely.

\begin{figure}[t]
	\centering
	
	\begin{subfigure}[t]{0.48\textwidth}
		\centering
		\includegraphics[width=\linewidth]{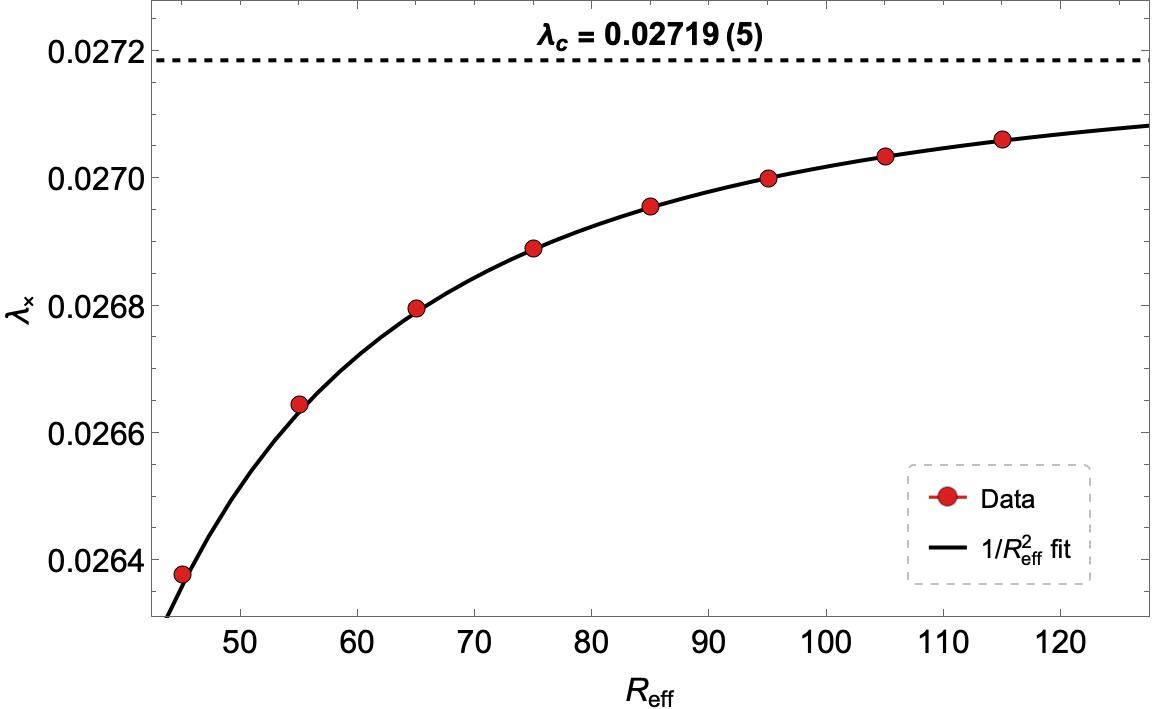}
		\caption{Large-$R$ extrapolation of Binder crossings.}
		\label{fig:crit-coupling}
	\end{subfigure}
	\hfill
	\begin{subfigure}[t]{0.48\textwidth}
		\centering
		\includegraphics[width=\linewidth]{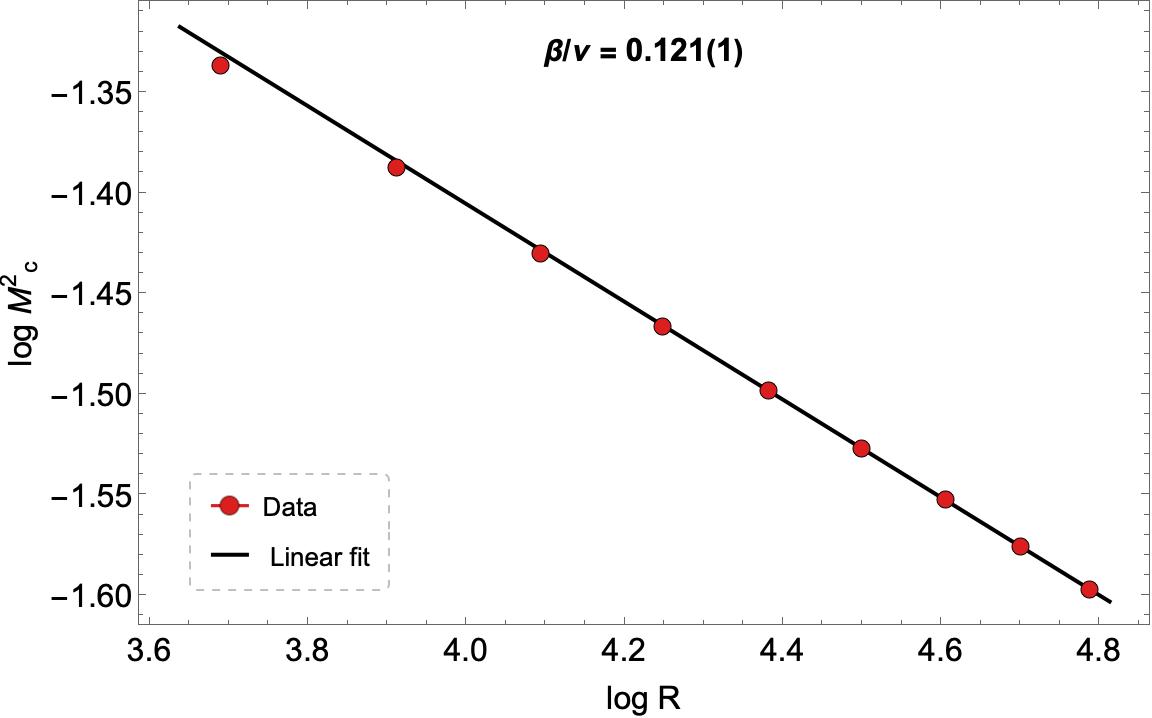}
		\caption{Scaling of order-parameter fluctuations $\langle M^2\rangle$.}
		\label{fig:beta-extraction}
	\end{subfigure}
	
	\vspace{0.4cm}
	
	\begin{subfigure}[t]{0.48\textwidth}
		\centering
		\includegraphics[width=\linewidth]{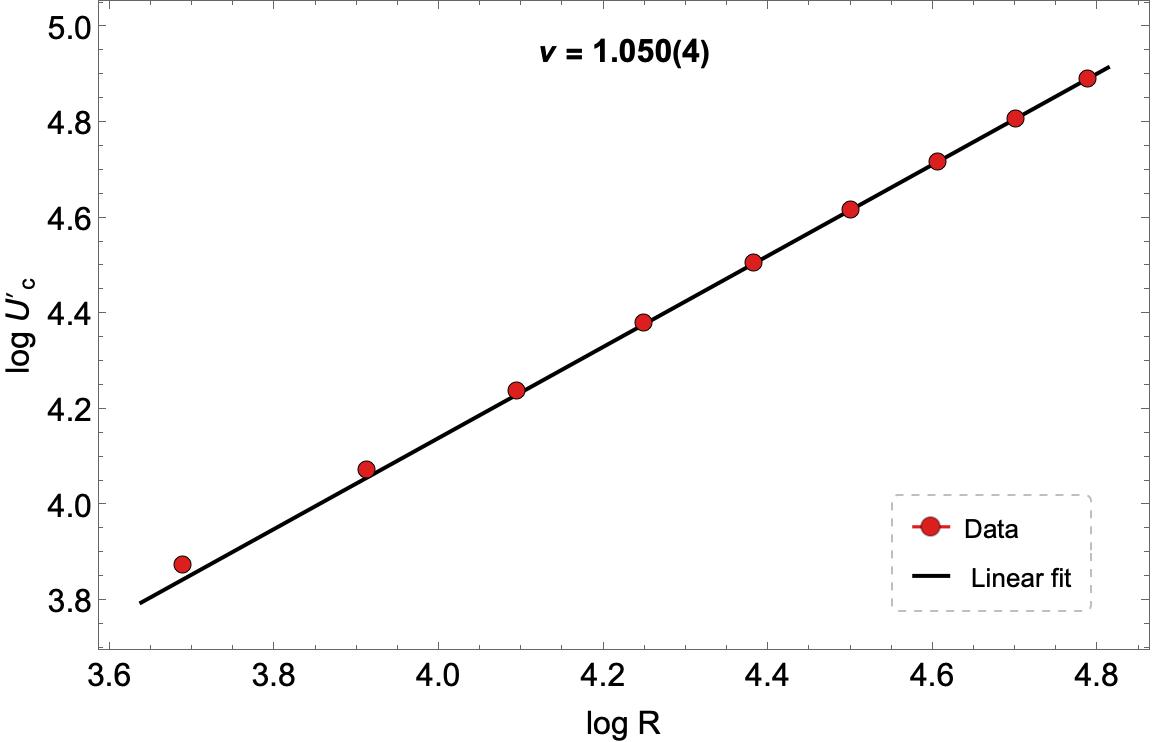}
		\caption{Scaling of Binder slopes.}
		\label{fig:nu-extraction}
	\end{subfigure}
	\hfill
	\begin{subfigure}[t]{0.48\textwidth}
		\centering
		\includegraphics[width=\linewidth]{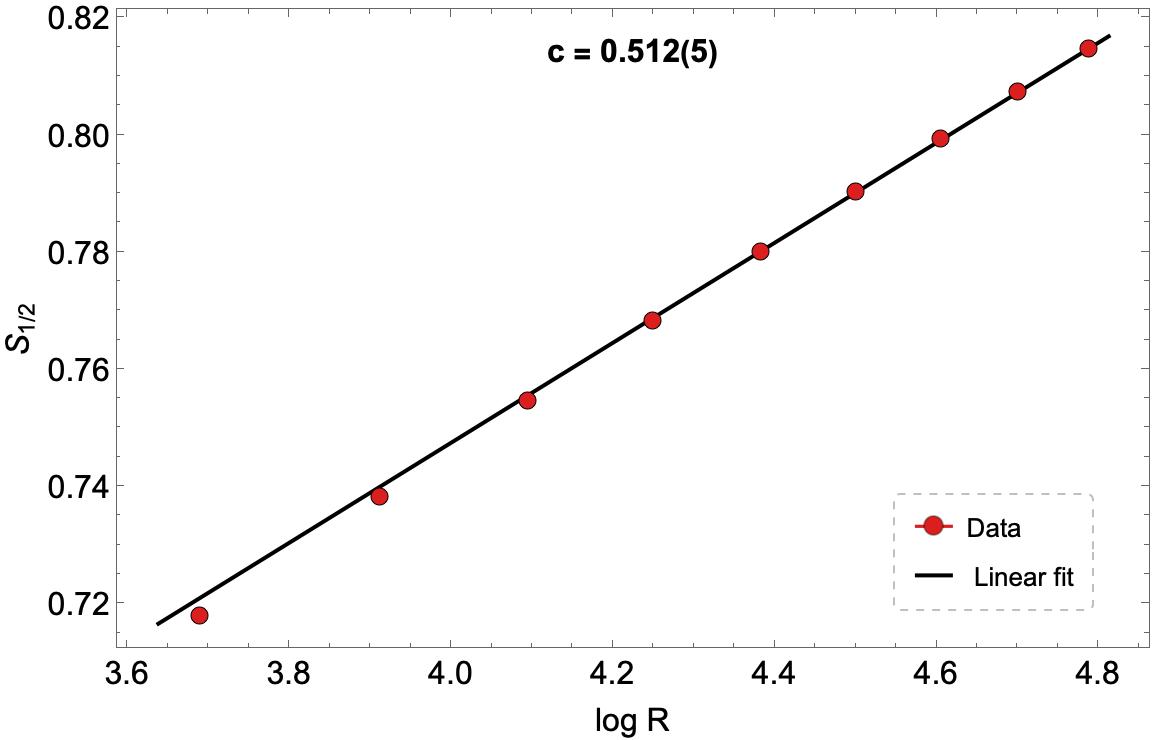}
		\caption{Scaling of half-chain entanglement entropy.}
		\label{fig:central-charge}
	\end{subfigure}
	
	\caption{
		 Critical data from finite-size scaling at the bulk critical point.
	}
	\label{fig:critical_data}
\end{figure}

For each pair of consecutive radii $(R_1,R_2)$, we
determine the crossing point $\lambda_\times(R_1,R_2)$ and associate it
with the effective radius
\begin{equation}
	R_{\rm eff}=\frac{R_1+R_2}{2}.
\end{equation}
The resulting sequence is shown in Fig.~\ref{fig:crit-coupling}. The
crossing points drift smoothly with $R_{\rm eff}$ and approach a
well-defined large-$R$ limit. Over the range of radii considered, this
drift is accurately described by
\begin{equation}
	\lambda_\times(R_{\rm eff})
	=
	\lambda_c+\frac{b}{R_{\rm eff}^2}.
	\label{eq:crossing_fit}
\end{equation}
At the fixed lattice spacing used here, the fit gives
\begin{equation}
	\lambda_c \simeq 0.02719(5).
	\label{eq:critical_lambda}
\end{equation}
The uncertainty in the extracted value comes from the variation under changes of the fitting range.  

Having determined $\lambda_c$, we next extract the critical exponents
from observables evaluated at the same coupling.  Let us first point out that the Binder crossings give a particularly stable determination of
$\lambda_c$. For a dimensionless scaling observable, the leading
critical value is independent of $R$. If the leading correction in
Eq.~\eqref{eq:binder_fss} scales as $R^{-\omega}$, the crossing shift
behaves parametrically as
$\lambda_\times-\lambda_c\sim R^{-(\omega+1/\nu)}$. The crossing
sequence can therefore converge faster than the observables we shall use to extract the critical exponents. The latter retain their ordinary
finite-$R$ corrections and need not yet lie fully in the asymptotic
scaling regime over the range of radii considered here.

First, we consider the scaling of the order-parameter fluctuations $\langle M^2\rangle$. Near the critical point, we expect
\begin{equation}
	M_c^2(R)
	\equiv
	\left.
	\langle M^2\rangle
	\right|_{\lambda=\lambda_c}
	\propto
	R^{-2\beta/\nu}.
	\label{eq:beta_scaling}
\end{equation}
The logarithmic form of this relation is shown in
Fig.~\ref{fig:beta-extraction}. The fitted slope gives
\begin{equation}
	\frac{\beta}{\nu}\simeq0.121(1),
	\label{eq:beta_nu_result}
\end{equation}
close to the two-dimensional Ising value
$(\beta/\nu)_{\rm Ising}=1/8$.

The correlation-length exponent $\nu$ can be obtained independently
from the Binder cumulant. Differentiating Eq.~\eqref{eq:binder_fss} at
the critical coupling gives, at leading order,
\begin{equation}
	U'_c(R)
	\equiv
	\left.
	\frac{\partial U}{\partial\lambda}
	\right|_{\lambda=\lambda_c}
	\propto
	R^{1/\nu}.
	\label{eq:nu_scaling}
\end{equation}
The fit shown in Fig.~\ref{fig:nu-extraction} yields
\begin{equation}
	\nu\simeq1.050(4),
	\label{eq:nu_result}
\end{equation}
again close to the Ising value $\nu_{\rm Ising}=1$. Combining the two
independently extracted exponents gives
\begin{equation}
	\beta\simeq0.127,
\end{equation}
close to $\beta_{\rm Ising}=1/8$.

As a further independent probe of the infrared fixed point, we extract
the central charge $c$ from the scaling of the entanglement entropy.
For a critical one-dimensional system with a boundary, the half-chain
entropy has the universal logarithmic form
\cite{calabrese2004EntanglementEntropy}
\begin{equation}
	S_{1/2}
	=
	\frac{c}{6}
	\log\!\left(\frac{R}{a_\rho}\right)
	+s_0+\cdots .
	\label{eq:entropy_scaling}
\end{equation}
Since $a_\rho$ is held fixed throughout the finite-size-scaling
sequence, it contributes only to the additive constant. The linear
dependence on $\log R$, shown in Fig.~\ref{fig:central-charge}, gives
\begin{equation}
	c\simeq0.512(5),
	\label{eq:c_result}
\end{equation}
close to the Ising value $c=1/2$.

The uncertainties quoted above include both the variation under changes
of the fitting range and the propagated uncertainty in $\lambda_c$.
They therefore reflect the numerical stability of the fits and
the uncertainty in locating the critical point, but do not include
possible subleading finite-$R$ corrections. The remaining deviations from the 2D Ising values are consistent with the accessible radii not yet being
fully in the asymptotic scaling regime.

\subsection{Conformal boundary condition at criticality}
\label{sec:critical_boundary_spectrum}
% -----------------------------------------------------------------------------

Having characterized the bulk critical point, we now return to the
second question posed at the beginning of this section: which conformal
boundary condition is reached at criticality? As discussed above, the
microscopic Dirichlet condition preserves the global $\mathbb Z_2$
symmetry and is therefore expected to flow to the symmetry-preserving
free/ordinary Ising boundary condition. The corresponding lowest
non-trivial global boundary dimensions are given in Eq.~\eqref{eq:ising_boundary}.

Having already located the bulk critical coupling from the Binder
analysis, we can now test this expectation directly from the low-lying global-AdS spectrum at that point. As in Sec.~\ref{sec:boundary_spectrum}, we extract the boundary
dimensions from the dimensionless energy gaps $
    \Delta_i(R,\lambda)
    =
    R\bigl[E_i(R,\lambda)-E_0(R,\lambda)\bigr]$.
At the bulk critical point,
\begin{equation}
    E_i(R,\lambda_c)-E_0(R,\lambda_c)
    =
    \frac{\Delta_i}{R}+\cdots ,
\end{equation}
so that the dimensionless gaps become independent of $R$ up to
finite-size corrections. Curves obtained at different radii are
therefore expected to cross in the vicinity of $\lambda_c$.

The behaviour away from criticality gives a simple interpretation of
this crossing. In the symmetric phase, the physical energy gap above the vacuum  approaches a constant as $R$ is increased, and so the dimensionless gap $\Delta_i(R)=R(E_i-E_0)$
grows with $R$. In the ordered phase, the lowest odd state becomes
degenerate with the ground state as the two broken vacua emerge, so
$\Delta_{\rm odd}$ instead decreases with $R$. At criticality,
$E_i-E_0\sim 1/R$, and the dimensionless gap approaches a constant.

\begin{figure}[h!]
    \centering
    \begin{subfigure}[t]{0.48\textwidth}
        \centering
        \includegraphics[width=\linewidth]{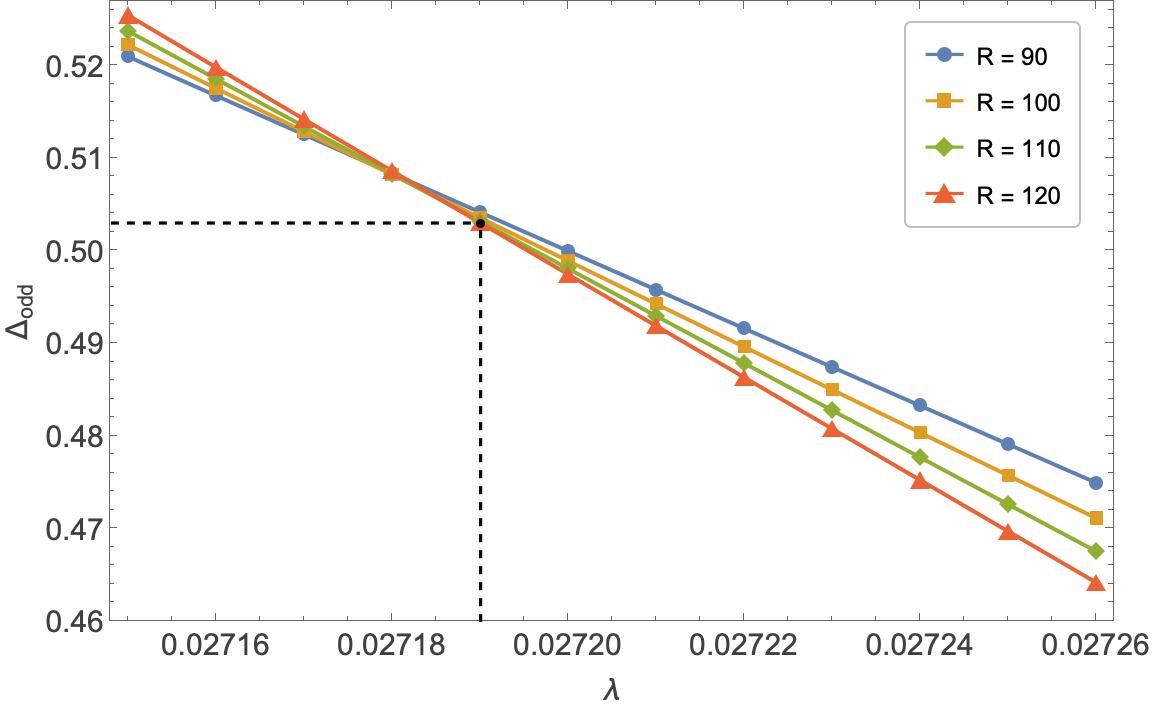}
        \caption{$\mathbb Z_2$-odd sector.}
        \label{fig:critical-odd-gap}
    \end{subfigure}
    \hfill
    \begin{subfigure}[t]{0.48\textwidth}
        \centering
        \includegraphics[width=\linewidth]{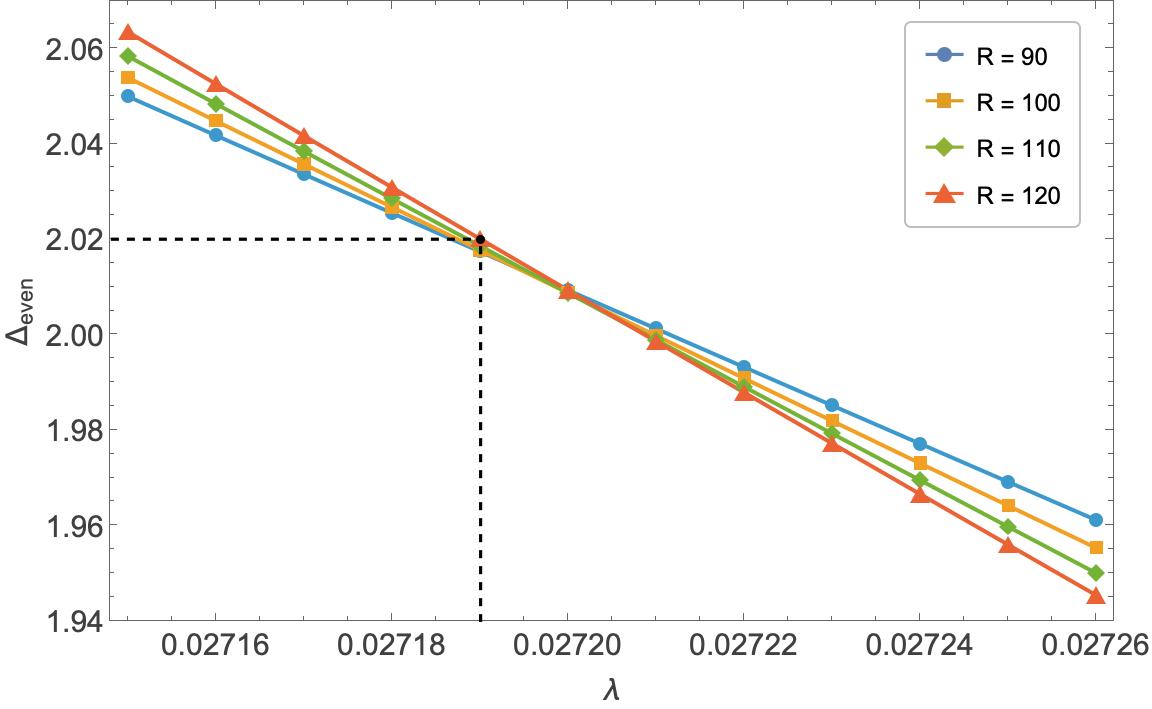}
        \caption{$\mathbb Z_2$-even sector.}
        \label{fig:critical-even-gap}
    \end{subfigure}

    \caption{
    Lowest non-trivial odd and even boundary gaps near the bulk
    critical point; dashed lines mark $\lambda_c$ and the corresponding
    $R=120$ values.
    }
    \label{fig:critical_boundary_spectrum}
\end{figure}

Since $\lambda_c$ is already known from the bulk analysis (Eq. \eqref{eq:critical_lambda}), we compute
the lowest non-trivial $\mathbb Z_2$-odd and even gaps only in a narrow
window around this value. The results are shown in
Fig.~\ref{fig:critical_boundary_spectrum} for
$R=90,100,110,$ and $120$. A clear crossing is visible in both parity
sectors, in the same critical region identified separately from the
Binder analysis. This also provides an independent consistency check.

The spectral crossings could in principle be used as an independent
way of locating the critical coupling. We do not pursue such an
extrapolation here. The excited-state gaps converge more slowly with
$R$ than the ground-state observables entering the Binder analysis,
and the available radii are not large enough for a comparably precise
large-$R$ determination. Instead, we evaluate the spectrum at the
critical coupling obtained from the Binder crossings, and quote the result at our largest
radius, $R=120$. We find
\begin{equation}
	\Delta_{\rm odd}(R=120,\lambda_c)
	\simeq 0.503(6),
	\qquad
	\Delta_{\rm even}(R=120,\lambda_c)
	\simeq 2.02(1) ,
	\label{eq:critical_boundary_results}
\end{equation}
with the corresponding values indicated by the dashed lines in
Fig.~\ref{fig:critical_boundary_spectrum}. The uncertainties quoted are solely from propagating the uncertainty in $\lambda_c$ and do not include residual finite-$R$ or radial-cutoff effects. In particular, the finite radial cutoff $\rho_{\max}=5R$ corresponds to a regulated endpoint $r_{\max}<\pi/2$, which can also produce small shifts in the extracted gaps. The numerical setup is summarized in Table~\ref{tab:critical_boundary_setup}.

\begin{table}[h!]
	\centering
	\caption{Numerical setup used for the critical boundary-spectrum calculations.}
	\label{tab:critical_boundary_setup}
	\begin{tabular}{ll}
		\toprule
		Renormalized mass &
		$m^2=0.01$ \\
		Spatial discretisation &
		Proper-distance lattice, Eq.~\ref{eq:proper_distance_grid} \\
		Lattice spacing &
		$a_\rho=1$ \\
		Radial cutoff &
		$\rho_{\max}=5R$ \\
		Local basis &
		Free-vacuum basis, Eq.~\ref{eq:free_vacuum_frequency} \\
		Local Hilbert-space dimension &
		$d=10$ \\
		Maximum bond dimension &
		$\chi=50$ \\
		\bottomrule
	\end{tabular}
\end{table}

Already at $R=120$, both gaps are in very good agreement with the
free/ordinary Ising values in
Eq.~\eqref{eq:ising_boundary}, with the small remaining deviations compatible with finite-$R$ and radial-cutoff effects. Together with the
preserved $\mathbb Z_2$ symmetry of the microscopic boundary
condition, the low-lying spectrum therefore identifies the conformal
boundary condition reached at the bulk critical point with the
free/ordinary Ising boundary condition.

This also provides a useful complement to the bulk analysis. The
Binder cumulant allows the critical point to be located efficiently
from ground-state observables, after which the global-AdS energy
spectrum at that coupling directly gives the associated boundary CFT
data. In this way the MPS construction gives access to both the bulk
critical behaviour and the conformal boundary spectrum without
requiring the excited-state spectrum itself to be used for precision
tuning of the critical coupling.

% -----------------------------------------------------------------------------
\section{Discussion and outlook}
\label{sec:discussion}
% -----------------------------------------------------------------------------

In this work, we developed a tensor-network
approach to study interacting QFTs in AdS$_2$. An important feature of the construction is that the low-lying states themselves are obtained
explicitly as matrix product states. Their energies determine the
boundary scaling dimensions, while the ground state can be used
independently to probe the bulk theory. Applying our approach to scalar $\phi^4$ theory, we followed the lowest odd and even boundary spectra  from weak coupling into the non-perturbative regime, and used finite-size scaling in the AdS radius to determine the bulk critical behaviour. The resulting critical data are consistent with the 2D Ising universality class,
while the low-lying spectrum at the same critical point identifies the symmetry-preserving free/ordinary Ising conformal boundary condition. The same setup therefore gives access to both the low-energy bulk states and the boundary spectrum.

The boundary spectra are only the first part of the boundary data that can be extracted. Since the corresponding states are
available explicitly, matrix elements of suitably renormalised operators can also be evaluated. After taking the boundary limit, these should give access to boundary correlation functions and OPE data as well.  We expect the main challenge in doing this to be numerical since such matrix elements are more sensitive than
energy gaps to lattice artefacts and extrapolation towards the AdS boundary. Related Hamiltonian approaches have nevertheless extracted form factors and spectral densities directly
from numerical eigenstates \cite{anand2017RGFlow,chen2022FormFactors},
while numerical finite-volume approaches have also demonstrated that OPE
coefficients can be recovered from microscopic states
\cite{hu2023OperatorProduct,lauchli2025ExactDiagonalization}.
A longer-term possibility would be to follow the resulting boundary
correlators towards large $R$ and connect them with flat-space scattering
data
\cite{paulos2017SmatrixBootstrap,carmi2019StudyQuantum,
	komatsu2020LandauDiagrams,vanrees2023QuantumField}.

The numerical set up presented here can also be extended to fermions.  Fermions are in some respects particularly
natural for tensor-network methods, since the local Hilbert
space at each site is finite-dimensional. This removes the additional occupation-number truncation needed
for bosons, although fermion discretisation and gauge constraints bring
their own complications. MPS calculations of the
finite-$N$ Gross--Neveu model in flat space provide one example
\cite{roose2020LatticeRegularisation}, while fermionic and non-Abelian
gauge theories in $1+1$ dimensions have also been studied with MPS
\cite{kuhn2015NonAbelianString ,banuls2017EfficientBasis}.
More recently, the Schwinger model has been treated directly in AdS$_2$
using tensor networks \cite{bharadwaj2026ConfinementScreening}.
The Gross--Neveu model would be a particularly interesting finite-$N$
application in AdS$_2$, complementing the existing large-$N$ results
\cite{carmi2019StudyQuantum}.

Extending the construction to AdS$_3$ is more demanding, since a
constant-time slice is two-dimensional and its spatial volume grows
rapidly towards the boundary. In proper radial coordinates the
circumference grows as $C(\rho)=2\pi R\sinh(\rho/R)$, making a uniform
proper-distance resolution increasingly expensive at large $\rho$.
For the low-energy quantities of interest here, however, such a uniform
resolution is not obviously necessary. Normalisable low-lying AdS states
are concentrated towards the interior, and at a bulk critical point the
curvature cuts off the flat-space correlation length at a scale of order
$R$; equivalently, a low-lying gap $E_1-E_0\sim\Delta_1/R$ corresponds to
a decay scale $\xi\sim R/\Delta_1$. The exponential growth of the spatial
volume is therefore still mild in the central region $\rho\lesssim R$.
In this work, the bulk scaling observables were measured
only inside $|\rho|<R$, even though the regulated geometry extended to
$\rho_{\max}=5R$. A discretisation with high resolution in the central
region and progressively coarser resolution outside it could therefore
be considerably more efficient than the uniform lattice used in this work. The remaining
2D many-body problem is of course substantially harder.
Nevertheless, $\phi^4$ theory in $2+1$ dimensions has already been studied
non-perturbatively using Hamiltonian-truncation methods
\cite{elias-miro2020ExploringHamiltonian,anand2021NonperturbativeDynamics},
while recent fuzzy-sphere calculations have used MPS to extract detailed
spectral and OPE data for the three-dimensional Ising and $O(2)$ CFTs
\cite{lauchli2025ExactDiagonalization,dey2026ConformalData}.
Together with existing lattice studies in AdS$_3$
\cite{brower2022HyperbolicLattice}, these results suggest that extending
the present approach beyond AdS$_2$ is worth exploring.

Several interesting questions already arise without leaving AdS$_2$.
The most direct continuation of this work would be to map
the finite-curvature phase diagram of $\phi^4$ theory itself. At fixed $R$, the theory depends on the two dimensionless couplings $m^2R^2$ and $\lambda R^2$, and this probes a different limit from the large-$R$ finite-size scaling regime. Determining the nature of the transition throughout this plane would provide a continuum finite-$N$ comparison with hyperbolic-lattice studies that find mean-field-like
critical behaviour \cite{iharagi2010PhaseTransition,gendiar2014MeanfieldUniversality, 	breuckmann2020CriticalProperties}, as well as field-theoretic analyses which reach different conclusions
about the finite-curvature fixed-point structure
\cite{benedetti2015CriticalBehavior,mnasri2015CriticalPhenomena}. It would also complement the existing large-$N$ and
Hamiltonian-truncation analyses directly in AdS$_2$
\cite{carmi2019StudyQuantum,hogervorst2021HamiltonianTruncation}.

A more substantial extension is the $O(2)$ model. At large $N$, the $O(N)$ model has a symmetry-breaking phase even in AdS$_2$, where curvature changes the infrared physics responsible for the usual flat-space obstruction \cite{carmi2019StudyQuantum}. It would be
interesting to determine what remains of this picture at $N=2$.
In flat 2D QFTs, spontaneous breaking of a
continuous internal symmetry is forbidden by Coleman's theorem
\cite{coleman1973ThereAre}, while the $O(2)$ model instead
exhibits the infrared physics associated with the
Berezinskii--Kosterlitz--Thouless transition
\cite{berezinskii1971DestructionLongrange,
	kosterlitz1973OrderingMetastability, grater1995KosterlitzThoulessPhase}.
The $O(2)$ rotor model has itself been studied successfully with MPS in
flat space \cite{milsted2016MatrixProduct}, whereas clock and XY models
on hyperbolic lattices indicate that negative curvature can qualitatively
change the BKT picture \cite{gendiar2008PhaseTransition}.
A continuum finite-$N$ calculation in AdS$_2$ could therefore ask directly
whether a curvature-supported ordered regime survives at $N=2$, and how
the flat-space infrared behaviour is recovered as $R\rightarrow\infty$.

Studying Sine--Gordon theory in AdS$_2$  provides another clean target. Its RG flow in AdS$_2$ has already been studied with the conformal bootstrap \cite{antunes2021BootstrappingRG}. The bootstrap constraints
are often saturated close to the UV fixed point and again in the
flat-space limit, but need not be saturated at intermediate values of the AdS scale. An MPS calculation could determine the actual finite-$R$ spectrum in precisely this region and, with the matrix elements discussed above, eventually provide boundary correlator data for direct comparison with the bootstrap bounds. Continuum sine--Gordon theory has already been studied
with large-scale MPS calculations in flat space, including the preparation and scattering of soliton excitations
\cite{calliari2025QuantumSimulating}.

Finally, the same setup could be used to vary the boundary physics itself. We kept a fixed microscopic Dirichlet condition in this work and identified the conformal boundary condition reached at bulk criticality. Adding boundary interactions would instead allow boundary RG flows between different boundary conditions to be followed directly through their low-lying spectra and matrix elements. One could also directly test the hypothesis in \cite{hogervorst2021HamiltonianTruncation} stating that all conformal boundary conditions for a given theory are connected via bulk RG flows.  Finally, this would also connect naturally with
recent proposals for monotonic quantities along RG flows generated by boundary perturbations \cite{bason2025FtheoremQuantum}.

\section*{Acknowledgements}
The author would like to thank Aninda Sinha, Xinan Zhou, Ujjwal Basumatary, Soumyadeep Chaudhuri
and Bernardo Zan for helpful discussions and comments on the manuscript.

\appendix

% =============================================================================
\section{Mass renormalization and tadpole evaluation in AdS}
\label{app:renormalization}
% =============================================================================

In this appendix, we describe the tadpole calculation entering the mass
renormalization prescription of Sec.~\ref{sec:phi4_renormalization}.
We first recall the free scalar mode expansion in global AdS$_2$ and
then give the corresponding lattice construction used in the numerical
calculations. Throughout this appendix, we set
$m_{\rm ref}^2=m^2$, as in all numerical results in the main text.

% -----------------------------------------------------------------------------
\subsection{Free scalar theory and the continuum tadpole}
% -----------------------------------------------------------------------------

The free scalar Hamiltonian is obtained from
Eq.~\eqref{eq:continuum_ads_hamiltonian} by taking
\[
\mathcal U(\phi)=\frac12 m^2\phi^2 ,
\]
so that
\begin{equation}
	H_0
	=
	\frac12
	\int_{-\pi/2}^{\pi/2}dr\,
	\left[
	\Pi^2
	+
	(\partial_r\phi)^2
	+
	\frac{R^2m^2}{\cos^2r}\phi^2
	\right].
	\label{eq:app_free_ads_hamiltonian}
\end{equation}
The free scaling dimension satisfies
\begin{equation}
	\Delta_\phi(\Delta_\phi-1)=m^2R^2,
	\qquad
	\Delta_\phi
	=
	\frac12+\sqrt{\frac14+m^2R^2}.
	\label{eq:app_delta_mass_relation}
\end{equation}

The normal modes obey
\begin{equation}
	\left[
	-\partial_r^2
	+
	\frac{m^2R^2}{\cos^2r}
	\right]u_n(r)
	=
	\omega_n^2 u_n(r),
	\qquad
	\omega_n=\Delta_\phi+n ,
	\qquad
	n=0,1,2,\ldots ,
	\label{eq:app_ads_mode_equation}
\end{equation}
and, with Dirichlet boundary conditions at
$r=\pm\pi/2$, may be chosen as
\begin{equation}
	u_n(r)
	=
	\mathcal N_n
	(\cos r)^{\Delta_\phi}
	C_n^{(\Delta_\phi)}(\sin r),
	\qquad
	\int_{-\pi/2}^{\pi/2}dr\,
	u_n(r)u_m(r)=\delta_{nm}.
	\label{eq:app_ads_modes}
\end{equation}

Expanding the field in these modes,
\begin{equation}
	\phi(t,r)
	=
	\sum_{n=0}^{\infty}
	\frac{u_n(r)}{\sqrt{2\omega_n}}
	\left(
	a_ne^{-i\omega_nt}
	+
	a_n^\dagger e^{i\omega_nt}
	\right),
	\qquad
	[a_n,a_m^\dagger]=\delta_{nm},
	\label{eq:app_field_mode_expansion}
\end{equation}
gives the equal-time two-point function
\begin{equation}
	G_{\rm AdS}(r,r';m^2)
	=
	\braket{\Omega_{\rm AdS}|
		\phi(r)\phi(r')
		|\Omega_{\rm AdS}}
	=
	\sum_{n=0}^{\infty}
	\frac{u_n(r)u_n(r')}{2\omega_n}.
	\label{eq:app_mode_sum_twopoint}
\end{equation}
The AdS tadpole is the coincident limit of this expression,
\begin{equation}
	Z_{\rm AdS}(m^2)
	=
	\lim_{r'\to r}
	G_{\rm AdS}(r,r';m^2),
	\label{eq:app_ads_tadpole_formal}
\end{equation}
which is ultraviolet divergent. In the numerical calculation we do
not evaluate this continuum coincident limit directly. Instead, the
same quantity is computed with the lattice regulator used for the
Hamiltonian, as we now describe.

% -----------------------------------------------------------------------------
\subsection{Tadpole on the lattice}
% -----------------------------------------------------------------------------

The regulated interval is represented by sites $r_j$, spacings
$\delta_j=r_{j+1}-r_j$, and quadrature weights $w_j$, as in
Sec.~\ref{sec:lattice_discretization}. With Dirichlet boundary
conditions, only the interior sites
\[
I=\{2,\ldots,N-1\}
\]
are dynamical. In terms of the canonically normalized variables of
Eq.~\eqref{eq:lattice_hamiltonian}, the free quadratic Hamiltonian is
\begin{equation}
	H_{0,\rm lat}
	=
	\frac12\sum_{j\in I}\widetilde\Pi_j^2
	+
	\frac12
	\sum_{i,j\in I}
	\widetilde\phi_i
	K^{\rm AdS}_{ij}
	\widetilde\phi_j .
	\label{eq:app_lattice_free_hamiltonian}
\end{equation}
The tridiagonal matrix $K^{\rm AdS}$ has diagonal entries
\begin{equation}
	K^{\rm AdS}_{jj}
	=
	R^2m^2\sec^2r_j
	+
	\frac{1}{w_j}
	\left(
	\frac{1}{\delta_{j-1}}
	+
	\frac{1}{\delta_j}
	\right),
	\qquad j\in I,
	\label{eq:app_K_ads_diag}
\end{equation}
and off-diagonal entries
\begin{equation}
	K^{\rm AdS}_{j,j+1}
	=
	K^{\rm AdS}_{j+1,j}
	=
	-\frac{1}{\delta_j\sqrt{w_jw_{j+1}}}.
	\label{eq:app_K_ads_offdiag}
\end{equation}
This form applies to both discretizations used in the main text; only
the site positions, spacings, and weights differ.

Diagonalizing the quadratic form,
\begin{equation}
	K^{\rm AdS}v^{(n)}
	=
	\Omega_n^2v^{(n)},
	\qquad
	\sum_{j\in I}
	v_j^{(n)}v_j^{(m)}
	=
	\delta_{nm},
	\label{eq:app_lattice_modes}
\end{equation}
gives the regulated AdS tadpole
\begin{equation}
	Z_{\rm AdS}^{(\mathrm{lat})}(r_j;m^2)
	=
	\braket{\phi_j^2}
	=
	\frac{1}{w_j}
	\sum_n
	\frac{\left(v_j^{(n)}\right)^2}{2\Omega_n}.
	\label{eq:app_lattice_ads_tadpole}
\end{equation}
At finite lattice spacing and radial cutoff this quantity is
position dependent because the regulator breaks exact AdS
homogeneity.

The flat-space reference is chosen to match the ultraviolet regulator
of the corresponding AdS calculation. The precise construction is
slightly different for the two discretizations used in the main text.

For the global-coordinate discretization, we use a flat lattice on a
strip with the same global-coordinate spacing and regulated interval.
Its physical length is
\begin{equation}
	L_{\rm flat}
	=
	2Rr_{\max}
	=
	(\pi-2\epsilon)R .
	\label{eq:app_flat_strip_length}
\end{equation}
The corresponding flat quadratic Hamiltonian is obtained by replacing
\begin{equation}
	R^2m^2\sec^2r_j
	\longrightarrow
	R^2m^2
	\label{eq:app_flat_replacement}
\end{equation}
in Eq.~\eqref{eq:app_K_ads_diag}, while keeping the same lattice
sites, spacings, quadrature weights, and Dirichlet boundary
conditions. The flat-strip tadpole is then obtained from the same
normal-mode construction as Eq.~\eqref{eq:app_lattice_ads_tadpole}. We use its
value at the centre of the strip and denote it by
$Z_{\rm flat}^{(\mathrm{lat})}(m^2)$.

For the proper-distance discretization used in the finite-size-scaling
analysis, the flat reference is instead defined on a uniform
flat-space lattice with the same fixed spacing $a_\rho$. Its spatial
extent is taken sufficiently large that the tadpole at the centre is
insensitive to the boundaries. Since $a_\rho$ is held fixed throughout
the scaling analysis, this defines the same flat-space lattice
renormalization scheme for every value of $R$.

In either case, we denote the corresponding flat tadpole by
$Z_{\rm flat}^{(\mathrm{lat})}(m^2)$ and define
\begin{equation}
	z_R^{(\mathrm{lat})}(m^2)
	=
	Z_{\rm AdS}^{(\mathrm{lat})}(0;m^2)
	-
	Z_{\rm flat}^{(\mathrm{lat})}(m^2).
	\label{eq:app_zR_lattice}
\end{equation}
This is the regulator-matched finite subtraction entering
Eq.~\eqref{eq:zR_definition}. For the global-coordinate calculation
the flat reference is therefore a finite strip, whereas for the
proper-distance calculation it is a large-volume flat lattice at
fixed $a_\rho$. The corresponding regulator limits and the final
mass counterterm are specified in
Sec.~\ref{sec:phi4_renormalization}.

% =============================================================================
\section{Perturbation theory for boundary spectrum}
\label{app:perturbation_theory}
% =============================================================================

In Sec.~\ref{sec:boundary_spectrum} we compared the low-lying
global-AdS gaps with perturbation theory at weak coupling. Since the perturbative curves in
Fig.~\ref{fig:ContMassGapOddEvenWeak} are used as numerical benchmarks
for the lattice calculation, we evaluate the perturbative corrections
directly in the same global-coordinate lattice regularization and
renormalization scheme used for the DMRG calculation.

Writing
\begin{equation}
	H_{\rm lat}
	=
	H_{0,\rm lat}
	+
	\bar\lambda\,H_{\rm lat}^{(1)},
	\qquad
	\bar\lambda=\lambda R^2 ,
	\label{eq:app_pert_expansion_hamiltonian}
\end{equation}
the interaction following from the renormalized Hamiltonian
Eq.~\eqref{eq:interaction_density_lattice} is
\begin{equation}
	H_{\rm lat}^{(1)}
	=
	\sum_{j\in I}
	\frac{\sec^2r_j}{w_j}\,
	\widetilde\phi_j^4
	+
	6
	\sum_{j\in I}
	\left[
	z_R(m^2)
	-
	Z_{\rm AdS}^{(\mathrm{lat})}(r_j;m^2)
	\right]
	\sec^2r_j\,
	\widetilde\phi_j^2 .
	\label{eq:lattice_pert_hamiltonian}
\end{equation}
Here $z_R(m^2)$ denotes the regulator-specific finite subtraction
appearing in Sec.~\ref{sec:phi4_renormalization}. For the
global-coordinate discretization it is obtained from
$z_R^{(\mathrm{lat})}$ by taking the limits in
Eq.~\eqref{eq:flat_corr_global}, using the flat-strip reference
described in Appendix~\ref{app:renormalization}.

Using the free lattice modes of
Eq.~\eqref{eq:app_lattice_modes}, we write
\begin{equation}
	\widetilde\phi_j
	=
	\sum_n
	\frac{v_j^{(n)}}{\sqrt{2\Omega_n}}
	\left(
	a_n+a_n^\dagger
	\right).
	\label{eq:app_lattice_field_modes}
\end{equation}
The free vacuum is annihilated by all $a_n$. The lowest odd and even
states are respectively
\begin{equation}
	|{\rm odd}\rangle
	=
	a_0^\dagger|0\rangle,
	\qquad
	|{\rm even}\rangle
	=
	\frac{1}{\sqrt2}
	\left(a_0^\dagger\right)^2|0\rangle ,
	\label{eq:app_free_odd_even_states}
\end{equation}
with free gaps
\begin{equation}
	\Delta_{\rm odd}^{(0)}=\Omega_0,
	\qquad
	\Delta_{\rm even}^{(0)}=2\Omega_0 .
	\label{eq:app_free_lattice_gaps}
\end{equation}
As the global-coordinate lattice is refined,
$\Omega_0\rightarrow\Delta_\phi$.

Normal-ordering Eq.~\eqref{eq:lattice_pert_hamiltonian} with respect
to the free lattice vacuum cancels the site-dependent AdS tadpole and
gives
\begin{equation}
	H_{\rm lat}^{(1)}
	=
	\sum_{j\in I}
	\frac{\sec^2r_j}{w_j}
	:\widetilde\phi_j^4:
	+
	6z_R(m^2)
	\sum_{j\in I}
	\sec^2r_j
	:\widetilde\phi_j^2:
	+
	\text{constant}.
	\label{eq:app_normal_ordered_lattice_interaction}
\end{equation}
The constant drops out of energy gaps.

We expand each gap as
\begin{equation}
	\Delta_\alpha
	=
	\Delta_\alpha^{(0)}
	+
	\bar\lambda\,\Delta_\alpha^{(1)}
	+
	\bar\lambda^2\,\Delta_\alpha^{(2)}
	+
	O(\bar\lambda^3),
	\qquad
	\alpha\in\{{\rm odd},{\rm even}\}.
	\label{eq:app_gap_expansion}
\end{equation}
At first order,
\begin{equation}
	\Delta_\alpha^{(1)}
	=
	\langle\alpha|
	H_{\rm lat}^{(1)}
	|\alpha\rangle
	-
	\langle0|
	H_{\rm lat}^{(1)}
	|0\rangle .
	\label{eq:app_first_order_gap}
\end{equation}
At second order, ordinary Rayleigh--Schr\"odinger perturbation theory
gives
\begin{equation}
	\Delta_\alpha^{(2)}
	=
	\sum_{\beta\neq\alpha}
	\frac{
		|\langle\beta|
		H_{\rm lat}^{(1)}
		|\alpha\rangle|^2
	}{
		E_\alpha^{(0)}-E_\beta^{(0)}
	}
	-
	\sum_{\beta\neq0}
	\frac{
		|\langle\beta|
		H_{\rm lat}^{(1)}
		|0\rangle|^2
	}{
		E_0^{(0)}-E_\beta^{(0)}
	} .
	\label{eq:second_order_gap}
\end{equation}
The second term subtracts the vacuum-energy correction because the
observable is an energy gap. By $\mathbb Z_2$ parity, the odd-gap sum
receives contributions only from odd intermediate states, while the
even-gap and vacuum sums involve only even states. All matrix elements
are evaluated algebraically in the free-mode Fock basis using
Eq.~\eqref{eq:app_lattice_field_modes}.

For the weak-coupling results in
Fig.~\ref{fig:ContMassGapOddEvenWeak} we use $m^2R^2=1$. The
perturbative benchmark was evaluated on the global-coordinate lattice
with $N=200$ and $\epsilon=10^{-3}$. Increasing $N$ and decreasing
$\epsilon$ changes the coefficients below by less than $10^{-3}$.
We obtain
\begin{align}
	\Delta_{\rm odd}
	&=
	1.617997
	-0.071894\,\bar\lambda
	-0.444241\,\bar\lambda^2
	+O(\bar\lambda^3),
	\label{eq:odd_gap_pert_numbers}
	\\
	\Delta_{\rm even}
	&=
	3.235994
	+0.626901\,\bar\lambda
	-2.040513\,\bar\lambda^2
	+O(\bar\lambda^3).
	\label{eq:even_gap_pert_numbers}
\end{align}
These are the second-order perturbative benchmarks shown as dashed
lines in Fig.~\ref{fig:ContMassGapOddEvenWeak}. Their agreement with
the DMRG gaps at weak coupling provides a direct check of the lattice
Hamiltonian, the renormalization prescription, and the excited-state
calculation.

% Bibliography commands can be added once the bibliography file is available:

\bibliographystyle{JHEP}
\bibliography{references}

\end{document}